\documentclass[11pt]{article}
\usepackage{amsmath}
\usepackage{amsfonts}
\usepackage{amsthm}
\newtheorem{thm}{Theorem}[section]
\newtheorem{lem}{Lemma}[section]

\theoremstyle{remark}
\newtheorem{rem}{Remark}
\numberwithin{equation}{section}
\newcommand \RR   {\mathbb{R}}
\def\be{\begin{equation}}
\def\ee{\end{equation}}
\def\v{\vec}

\def\b{\bullet}
\def\la{\label}

\def\pa{\partial}
\def\f{\frac}

\def\t{\tilde}

\def\g{\gamma}

\def\r{\rho}
\def\b{\bar}
\def\f{\frac}

\def\t{\tau}

\begin{document}
\title{ Existence and Nonlinear Stability of Rotating Star Solutions of the Compressible Euler-Poisson Equations }
\author{ Tao Luo \&  Joel Smoller}
\date{}
\maketitle
\begin{abstract}We prove existence of rotating star solutions which are steady-state solutions of the compressible isentropic Euler-Poisson (EP)
equations in 3 spatial dimensions, with prescribed angular momentum and total mass. This problem can be formulated as a variational problem of
finding a minimizer of an energy functional in a broader class  of functions having less symmetry than those functions considered in the classical
Auchmuty-Beals paper. We prove the nonlinear dynamical stability of these solutions with  perturbations having the same total mass and symmetry as the rotating
star solution. We also prove local in time stability of $W^{1, \infty}(\RR^3)$ solutions where the perturbations are entropy-weak solutions of the
EP equations. Finally, we give a uniform (in time) a-priori estimate for entropy-weak solutions of the EP equations.\end{abstract}
\tableofcontents

\section{Introduction}

The motion of a
 compressible isentropic perfect fluid with  self-gravitation is
modeled by the  Euler-Poisson equations in three space dimensions  (cf \cite{ch}):
\begin{equation}\label{1.1}
\begin{cases}
&\rho_t+\nabla\cdot(\rho {\bf v})=0,\\
&(\rho {{\bf v}})_t+\nabla\cdot(\rho {\bf v}\otimes {\bf v})+\nabla p(\rho)=-\rho \nabla \Phi,\\
&\Delta \Phi=4\pi \rho.
\end{cases}
\end{equation}
Here $\rho$, ${\bf v}=(v_1, v_2, v_3)$, $p(\rho)$ and $\Phi$ denote
the density, velocity, pressure and gravitational
potential, respectively.
  The gravitational potential is given by
\be\label{phi}\Phi(x)=-\int_{\RR^3} \frac{\rho(y)}{|x-y|}dy =-\rho\ast \frac{1}{|x|},\ee where $\ast$ denotes convolution.  The momentum $\rho{\bf v}$ is denoted by ${\bf m}=(m_1, m_2, m_3)$.  System (\ref{1.1}) is used to model the evolution of a Newtonian gaseous star (\cite {ch}). In the  study of time-independent solutions of system (\ref{1.1}), there are two important cases, non-rotating stars and rotating stars.  A non-rotating star solution
is a time-independent spherical symmetric solution of the form $(\rho_N, 0, \Phi_N)(x)$ (the velocity is zero), with
$\Phi_ N(x)=-\rho_N\ast \frac{1}{|x|}$. A rotating star solution models a star rotating around the $x_3$-axis ($x=(x_1,\ x_2,\ x_3))$ with prescribed angular momentum (per unit mass),  or  angular velocity.  The existence and properties of stationary non-rotating star solutions is classical (cf. \cite{ch}).
In contrast, the study for rotating stars is more challenging and of  significance in both astrophysics and
mathematics.
A rigorous mathematical theory for rotating stars of compressible fluids was initiated by  Auchmuty \& Beals (\cite {AB}) in 1971.  The existence and properties of rotating star solutions were obtained by Auchmuty \& Beals (\cite{AB}), Auchmuty(\cite{Au}), Caffarelli \& Friedman (\cite{CF}), Friedman \&  Turkington(\cite{FT1},  \cite{FT2}),  Li(\cite{Li1}), Chanillo \& Li(\cite{Li2}), and  Luo \& Smoller (\cite{LS}). In \cite{mc}, McCann proved an existence result for rotating binary stars.

The existence of rotating star solutions of compressible fluids was first obtained by  Auchmuty \& Beals (\cite {AB}) who formulated this problem as a variational problem of finding a minimizer of the energy
functional $F(\r)$, (which will be defined in Section 2), in the class of functions $W_{M, S}=W_M\cap W_S$, where
 $W_M$ is the set of integrable functions  $\r: \RR^3\to \RR^+$ which are a.e. non-negative, axi-symmetric, of total
mass $M=\int_{\RR^3}\r(x)dx$,  and having a finite rotational kinetic energy (precise statements can be found in Section 2). $W_S$ is defined by  \begin{equation}\label{W2'}
  W_S=\{\r: \RR^3\to \RR^+,\   \r(x_1, x_2, -x_3)=\r(x_1, x_2, x_3),\ x_i\in \RR,\ i=1, 2,  3\}.
  \end{equation}
In this paper, we first give a proof of the existence of a minimizer of the energy functional $F(\r)$ in the wider class of functions $W_M$. Our proof is quite different from that in \cite{AB}. As in \cite{AB},  the main difficulty in the proof is  the loss of compactness due to
the unboundedness of  $\RR^3$. The method in \cite {AB} is to minimize the functional $F$ on $W_R=\{\r\in W_{M,S}, \r(x)=0\,  \ |x|>R\}$ and to obtain some uniform
estimates on the support of the minimizer. Our method is to use the concentration-compactness method due to P. L. Lions (\cite{lions}), which was also
used in \cite{rein1} to prove the existence of non-rotating star solutions.  The reason that we seek  minimizers in $W_M$ instead of $W_{M, S}$ is that we want to discuss the full stability problem dynamically in a more general context with less restrictions on the symmetry of solutions.

  The dynamical  stability of these steady-state solutions is an important question. The linearized stability and instability  for non-rotating stars and rotating stars  were   discussed by Lin (\cite{lin} ), Lebovitz (\cite{lebovitz})  and Lebovitz \& Lifschitz (\cite{lebovitz1}). The nonlinear dynamical stability of  {\it non-rotating}  star solutions was studied by Rein (\cite {Rein}) via  an energy-Casimir technique. It should mentioned here that the energy-Casimir technique was used in \cite{guo1} to study the stability problem in stellar dynamics.   Roughly speaking, for $p(\rho)=\rho^{\gamma}$, the result in \cite {Rein} says that if the initial data of
the Euler-Poisson equations (\ref{1.1}) is close to the non-rotating star solution in some topology, then the solution of (\ref{1.1}) with the same
total mass as the non-rotating star,  stays close
to the non-rotating solution in the same topology as long as the solution preserves both the energy $E(t)$ which is
defined by
\begin{equation}\label{energy1}
E(t)=\int_{\RR^3}\left(\frac{p(\rho)}{\gamma-1}+\frac{1}{2}\r|{\bf v}|^2\right)(x, t)dx-\frac{1}{8\pi}\int_{\RR^3}|\nabla \Phi|^2(x, t)dx,\end{equation}
and the total mass $\int_{\RR^3}\rho(x, t)dx$.  An interesting feature of the energy is that it has both positive
and  negative parts, making the analysis difficult. For  solutions of (\ref{1.1}) without shock waves, energy is conserved. For solutions with shock waves,  the energy $E(t)$ is non-increasing due to the entropy condition associated with shock waves (cf. \cite{lax} and \cite{smoller}).
 In this  paper we  extend the above  nonlinear stability results   to  {\it rotating} stars.

As in the  non-rotating star case (\cite{Rein}), our nonlinear stability result is in the class of   solutions having the same
total mass as that of the rotating steady-state solution. For  solutions with different total masses,  we investigate the nonlinear dynamical stability of a solution $\bar u=(\bar \rho, \bar {\bf v}, \bar \Phi)\in W^{1, \infty}_{loc}$, (which includes both rotating and non-rotating stars), in the context of weak entropy solutions, for more general perturbations not necessarily having the same mass as $\bar u$,  under
some assumptions on the $L^{\infty}$-norm and the support of the solutions.  This is achieved by using the techniques of relative entropies together with a careful
analysis of the gravitational energy; i.e., the negative part in the total energy $E(t)$. It should be mentioned here that the method of relative entropies was
used by Dafermos (\cite{dafermos}) and Chen/Frid {\cite{chen}) to study the stability and behavior of solutions of hyperbolic conservation laws. The main difficulty in applying this method to the the Euler-Poisson equations (\ref{1.1}) is again due to the non-definiteness of the  energy density.    We also give a uniform a priori estimate for the  weak solutions of Cauchy problem of  (\ref{1.1})  satisfying the entropy conditions.

This paper is organized as follows: in Section 2, we prove the existence of rotating star solutions which are the minimizers of an energy functional $F$ in $W_M$  with  prescribed total mass and
angular momentum with finite rotational kinetic energy. We also derive  some properties concerning
 the minimizing sequence.  These properties are interesting,  and are important for our stability analysis.  In Section 3, we prove our nonlinear stability result for rotating stars. Section 4 is devoted to the stability result
for the entropy weak solutions and in Section 5, we obtain uniform in time a priori estimates for entropy weak solutions.

Throughout this paper, for simplicity of presentation, we assume that the pressure function $p(\rho)$ satisfies the usual $\g$-law,
\be p(\rho)=\rho^{\g},\  \rho\ge 0, \ee
for some  $\g>1$.  We now introduce some notation which will be used throughout this paper. We use $\int$ to denote $\int_{\RR^3}$, and use $||\cdot||_q$ to denote $||\cdot||_{L^q(\RR^3)}$.  For any point $x=(x_1, x_2, x_3)\in \RR^3$, let
\be\label{1.5'} r(x)=\sqrt{x_1^2+x_2^2}, \  z(x)=x_3,\  B_R(x)=\{y\in \RR^3, \ |y-x|<R\}.\ee
For any function $f\in L^{1}(\RR^3)$, we define the operator $B$ by
\be\label{B} B f(x)=\int \frac{f(y)}{|x-y|}dy =f\ast \frac{1}{|x|}.\ee
Also, we use $\nabla$ to denote the spatial gradient, i.e.,  $\nabla=\nabla_x=(\pa_{x_1},\ \pa_{x_2}, \ \pa_{x_3})$.  $C$ will denote a generic positive constant.

\section{Existence of Rotating Star Solutions}
A rotating star solution $(\tilde \rho, \tilde {\bf v},\tilde \Phi)(r, z)$, where $r=\sqrt{x_1^2+x_2^2}$  and $z=x_3$, $x=(x_1, x_2, x_3)\in \RR^3,$  is an {\it axi-symmetric}  time-independent solution of system (\ref{1.1}), which models a star rotating about the $x_3$-axis. Suppose the angular momentum (per unit mass),  $J(m_{\tilde \r}(r))$ is prescribed, where
\be m_{\tilde \r}(r)=\int_{\sqrt{x_1^2+x_2^2}<r}\tilde \r(x)dx=\int_0^r 2\pi s\int_{-\infty}^{+\infty}\tilde \r(s, z)dsdz, \ee
is the mass in the cylinder $\{x=(x_1, x_2, x_3): \sqrt{x_1^2+x_2^2}<r\}$, and $J$ is a given function.
   In this case, the velocity field
   $\tilde {\bf v}(x)=(v_1, v_2, v_3)$ takes the form $$\tilde {\bf v}(x)=(-\f{x_2 J(m_{\tilde \r}(r))}{r^2},  \f{x_1 J(m_{\tilde \r}(r))}{r^2}, 0). $$ Substituting  this in (\ref{1.1}), we find that $\tilde \r(r, z)$ satisfies the following two equations:
   \be\label{03}\begin{cases} &\partial_r p(\tilde \r)=\tilde \r\pa_r (B \tilde \r)+\tilde \r L(m_{\tilde \r}(r)r^{-3}, \\
                     &\partial_z p(\tilde \r)=\tilde \r\pa_z (B \tilde \r),
\end{cases}\ee
where the operator $B$ is defined in (\ref{B}), and
 $$ L(m_{\tilde \r})=J^2(m_{\tilde \r})$$ is the square of the angular momentum. For any function $\r\ge 0$ and $\g>1$,
we define
\be\label{z1}
A(\r)=\f{p(\r)}{\g-1}=\f{\r^\g}{\g-1}.\ee
  It is easy to verify that (cf. \cite{AB}) (\ref{03}) is equivalent to
  \be\label{z2}
  A'(\tilde \r(x))+\int_{r(x)}^{\infty}L(m_{\tilde \r}(s)s^{-3}ds-B\tilde \r(x)=\lambda, \qquad {\rm where~} \tilde \r(x)>0,\ee
  for some constant $\lambda$.   Here $r(x)$ and  $z(x)$ are as in (\ref{1.5'}).
  In \cite{AB},   Auchmuty and Beals  formulated the problem of finding solutions of (\ref{z2})  as the following  variational problem.
First, let $M$ be a positive constant and let $W_M$ be the set of functions $\r$ defined by (cf. (1.5)),
\begin{align*}W_M=&\{\r: \RR^3\to \RR,\ \r  {\rm~is~axisymmetric, ~}\rho\ge 0, a.e.,\ \rho\in L^1(\RR^3)\cap L^{\gamma}(\RR^3),\\
 &\int\rho(x)dx=M, \ \int\f{\rho(x)L(m_{\r}(r(x)))}{r(x)^2}dx<+\infty.\}\end{align*}

 For $\rho\in W_M$,  we define the {\bf energy functional} $F$ by
  \begin{align}\label{E}
  F(\rho)&=\int
  [A(\rho(x))+\f{1}{2}\f{\rho(x)L(m_{\r}(r(x)))}{r(x)^2}-\frac{1}{2}\rho(x)\cdot B\rho(x)]dx\notag\\
  &=\int
  [A(\rho(x))+\f{1}{2}\f{\rho(x)L(m_{\r}(r(x)))}{r(x)^2}]dx-\frac{1}{8\pi}||\nabla B\rho||_2^2.
  \end{align}
 ($\frac{1}{8\pi}|| B\rho||_2^2<+\infty$ follows from $\rho\in L^1(\RR^3)\cap L^{\gamma}(\RR^3)$ and  Lemma 2.3 if $\g\ge 4/3$.) In (\ref{E}),  the first
 term denotes the potential energy, the middle term denotes the rotational kinetic energy and the third term is the gravitational energy.
 Assume that the function  $L\in C^1[0, M]$ and satisfies
  \be\label{L} \ L(0)=0,\ L(m)\ge 0, \ for~ 0\le m\le M.\ee
Auchmuty and Beals (cf. \cite{AB}) proved the existence of a minimizer of the
functional $F(\rho)$ in the class of functions  $W_{M, S}=W_M\cap W_S$, where
  \begin{equation}\label{W2}
  W_S=\{ \r: \RR^3\to \RR,\   \r(x_1, x_2, -x_3)=\r(x_1, x_2, x_3),\ x_i\in \RR, i=1,\ 2, \ 3\}.
  \end{equation} Their result is given  in the following theorem.
\begin{thm}\label{aa1}(\cite{AB}).  If $\gamma>4/3$ and (\ref{L}) holds,  then there exists a function $\hat \r(x)\in W_{M, S}$ which minimizes $F(\rho)$ in  $W_{M, S}$. Moreover, if
\be\label{G}
G=\{x\in \RR^3:\  \hat \r(x)>0\},\ee
Then $\bar G$ is a compact set in $\RR^3$, and $\hat \r\in C^1(G)\cap C^{\beta}(\RR^3)$ for some $0<\beta<1$. Furthermore, there exists a constant $\mu<0$
such that
\be\label{lambda}
\begin{cases}
& A'(\hat \r(x))+\int_{r(x)}^{\infty}L(m_{\hat \r}(s)s^{-3}ds-B\hat \r(x)=\mu, \qquad x\in G,\\
&\int_{r(x)}^{\infty}L(m_{\hat \r}(s)s^{-3}ds-B\hat \r(x)\ge \mu, \qquad x\in \RR^3-G.\end{cases}\ee
\end{thm}
In this paper, we are interested in the  minimizer of functional $F$ in the {\it larger} class  $W_M$. By the same argument as in \cite{AB}, it is easy to prove the following theorem on the regularity of the minimizer.
\begin{thm}\label{ab} Let $\tilde \r$ be a minimizer of the energy functional $F$ in $W_M$ and let \be\label{G1}
\Gamma=\{x\in \RR^3:\  \tilde \r(x)>0\}.\ee If $\g>6/5$, then $\tilde \r\in C(\RR^3)\cap C^1(\Gamma)$.
Moreover,
 there exists a constant $\lambda$
such that
\be\label{lambda1}
\begin{cases}
& A'(\tilde \r(x))+\int_{r(x)}^{\infty}L(m_{\tilde \r}(s)s^{-3}ds-B\tilde \r(x)=\lambda, \qquad x\in \Gamma,\\
&\int_{r(x)}^{\infty}L(m_{\tilde \r}(s)s^{-3}ds-B\tilde \r(x)\ge \lambda, \qquad x\in \RR^3-\Gamma.\end{cases}\ee
\end{thm}
We call such a minimizer $\tilde \r$ a {\it rotating star} solution with  total mass $M$ and angular momentum
$\sqrt{ L(m)}$.

 In this paper, we prove the existence of a minimizer for the  functional $F$ in the class $W_M$. For this purpose, in addition to (\ref{L}),
we require that $L$ satisfies the following conditions:
\be\label{L1} L(a m)\ge a^{4/3}L(m), \ 0<a\le 1,\ 0\le m\le M,\ee
\be\la{L2'} L'(m)\ge 0,\qquad 0\le m\le M.\ee
\begin{rem} Condition (\ref{L2'}) is called the S$\ddot{\rm o}$lberg stability criterion, see [33,  Section 7.3]. This condition was also used by Auchmuty in \cite{Au}
for the study of global branching of rotating star solutions.\end{rem}
Our main result in this section is the following theorem.

\begin{thm}\label{aa}  Suppose  that $\gamma>4/3$ and the square of the angular momentum  $L$ satisfies (\ref{L}), (\ref{L1}) and (\ref{L2'}).  Then
the following hold:\\
(1) the functional $F$ is bounded below on $W_M$ and $\inf_{W_M} F(\rho)<0$,\\
(2) if $\{\r^i\}\subset W_M $ is a minimizing sequence for the functional $F$,
then there exist a sequence of vertical shifts $a_i{\bf e_3}$ ($a_i\in \RR$, ${\bf e_3}=(0, 0, 1)$),  a subsequence of $\{\r^i\}$,  (still labeled  $\{\r^i\}$),  and a function $\tilde \r\in W_M$, such that for any $\epsilon>0$ there exists $R>0$ with
\be\label{2.15}
\int_{a_i{\bf e_3}+B_R(0)}\r^i(x)dx\ge M-\epsilon, \quad i\in \mathbb{N},\ee
and
\be\label{2.16} T\r^i(x):=\r^i(x+a_i{\bf e_3})\rightharpoonup \tilde \r,\  weakly~in~L^{\g}(\RR^3),\ as\  i\to \infty.\ee

\noindent Moreover
(3)
\be\label{2.17}\nabla B (T\rho^i)\to \nabla B(\tilde \r)~ strongly~ in~L^2(\RR^3),\ as\  i\to \infty. \ee
 (4) $\tilde \r$ is a minimizer of $F$ in $W_M$.

\end{thm}

Thus $\tilde \r$ is a rotating star solution with  total mass $M$ and angular momentum
$\sqrt {L}$.

\begin{rem} It is easy to verify  that the functional $F$ is invariant under any vertical shift, i.e., if $\r(\cdot)\in W_M$, then
$\bar \r(x)=:\r(x+a{\bf e_3})\in W_M$ and $F(\bar \r)=F(\r)$ for any $a\in \RR$. Therefore, if $\{\r^i\}$ is a minimizing sequence of $F$ in
$W_M$, then $\{T\r^i\}$ defined in (2.15) is also a minimizing sequence in $W_M$.
\end{rem}

\begin{rem} In \cite{FT1}, \cite{FT2} and \cite{Li2}, the diameter estimate of rotating star solutions with the symmetry $\tilde \r(r,-z)=
\tilde \r(r, z)$ was obtained.  The ideas and techniques developed in \cite{FT1}, \cite{FT2} and \cite{Li2} should also be applied to obtain the diameter estimates for  the rotating star solutions in Theorem \ref{aa}.
Due to the length of this paper, we leave this issue for the future study.
\end{rem}
 Theorem \ref{aa} is proved in a sequence of lemmas.  We first give some  inequalities which will be used later.  We begin with   Young's inequality (see \cite{GT}, p. 146.)
 \begin{lem} If $f\in L^p\cap L^r$, $1\le p<q<r\le +\infty$, then
\be\label{young} ||f||_q\le ||f||_p^a||f||_r^{1-a}, \qquad a=\f{q^{-1}-r^{-1}}{p^{-1}-r^{-1}}.\ee \end{lem}
The following two lemmas are proved in \cite{AB}.
\begin{lem}\label{bf1'} Suppose the function $f\in L^1(\RR^3)\cap L^{q}(\RR^3)$. If $1<q\le 3/2$, then $Bf=:f\ast\f{1}{|x|}$ is in $L^{r}(\RR^3)$ for
$3<r<3q/(3-2q)$, and
\be\label{bf1} || Bf||_r\le C \left(||f||_1^b||f||_q^{1-b}+||f||_1^c||f||_q^{1-c}\right),\ee for some constants $C>0$, $0<b<1$,  and $0<c<1$.
If $q>3/2$, then $Bf(x)$ is a bounded continuous function, and  satisfies (2.18) with $r=\infty.$
\end{lem}
\begin{lem}\label{lem2.2} For any function $f\in L^1(\RR^3)\cap L^{\gamma}(\RR^3)$, if $\gamma\ge 4/3$,  then $\nabla Bf\in L^2(\RR^3)$. Moreover,
\be\label{bf2} |\int f(x)Bf(x)dx|=\f{1}{4\pi}||\nabla Bf||_2^2\le C \left(\int|f|^{4/3}(x)dx\right)\left(\int|f|(x)dx\right)^{2/3},\ee for some constant $C$.
\end{lem}
Throughout this paper, we assume the function $L$,  the square of the angular momentum satisfies conditions (\ref{L}),  (\ref{L1})and (2.13). Let
\be\label{fm}
f_M=\inf_{\r\in W_M}F(\r).\ee We begin our analysis with the following lemma.
\begin{lem}\label{lem4.2} Suppose $\g>4/3$. If $\r\in W_M$, then there exist two positive constants $C_1$ and $C_2$ depending only on $\g$ and $M$ such that
\be\label{us} \int [\r^{\g}(x)+\f{\rho(x)L(m_{\r}(r(x)))}{r(x)^2}] dx\le C_1 F(\r)+C_2.\ee
This implies $$f_M>-\infty, $$
where $f_M$ is defined in (2.20). \end{lem}
\begin{proof}
Using (\ref{bf2}), we have, for $\r\in W_M$,
\begin{align}\label{00}
F(\r)&=\int
  [A(\rho)+\f{1}{2}\f{\rho(x)L(m_{\r}(r(x)))}{r(x)^2}-\frac{1}{2}\rho B\rho]dx\notag\\
  &\ge \int
  [A(\rho)+\f{1}{2}\f{\rho(x)L(m_{\r}(r(x)))}{r(x)^2}]dx -C\int \r^{4/3}dx(\int \r dx)^{2/3}\notag\\
  &=\int
  [A(\rho)+\f{1}{2}\f{\rho(x)L(m_{\r}(r(x)))}{r(x)^2}]dx-CM^{2/3}\int \r^{4/3}dx.
  \end{align}
   Taking $p=1$, $q=4/3$, $r=\g$,  and $a=\f{\f{3}{4}\g-1}{\g-1}$ in Young's inequality (\ref{young}), we obtain,
  \be ||\r||_{4/3}\le ||\r||_1^a||\r||_{\g}^{1-a}=M^a||\r||_{\g}^{1-a}.\ee
  This is
  \be\label{haha}\int\r^{4/3}dx\le M^{\f{4}{3}a}(\int\r^\g dx)^b,
  \ee
  where $b=\f{1}{3(\g-1)}$. Since $\g>4/3$, we have $0<b<1$. Therefore, (\ref{00}) and (\ref{haha}) imply
  \be\la{0001}  \int
  [A(\rho)+\f{1}{2}\f{\rho(x)L(m_{\r}(r(x)))}{r(x)^2}]dx\le F(\r)+C(\g-1)^bM^{\f{4}{3}a+\f{2}{3}}(\int A(\r)dx)^b.\ee
  Using (\ref{haha}) and the  inequality (cf. \cite{GT} p. 145)
  \be\la{0002}\alpha\beta\le \epsilon\alpha^s+\epsilon^{-t/s}\beta^t,\ee
   if  $s^{-1}+t^{-1}=1$ ($s, t>1$) and $\epsilon>0$, since $b<1$,   we can bound the last term in (\ref{0001}) by $\f{1}{2}\int A(\rho)dx+C_2$,
   where $C_2$ is a constant depending only on $M$ and $\g$ (we can take $\epsilon=1/2$ and $s=1/b$ and $t=(1-s^{-1})^{-1}$ in (\ref{0002}) since $s>1$ due to $0<b<1$). This  implies (\ref{us}).\end{proof}
We also need the following lemma.
\begin{lem}\label{lem4.4} Suppose $\g>4/3$. Then \\
(a) $f_M<0$ for every $M>0$,\\
(b) if (\ref{L1}) holds, then $f_{\bar M}\ge (\bar M/M)^{5/3}f_M$  for every $M>\bar M>0$ .\end{lem}
\begin{proof}
It follows from \cite{AB} that there exists  $\hat \r\in W_{M, S}\subset W_M$ such that $F(\hat \r)=\inf_{\r\in W_{M, S}}F(\r)$. By Theorem \ref{aa1}, it is easy to verify that
the triple $(\hat \r, \hat {\bf v}, \hat\Phi)$ is a time-independent solution of the Euler-Poisson equations (\ref{1.1}) in the region $G=\{x\in \RR^3:\  \hat \r(x)>0\},$  where $\hat {\bf v}=(-\f{x_2 J(m_{\hat \r}(r))}{r},  \f{x_1 J(m_{\hat \r}(r))}{r}, 0)$ and $\hat \Phi=-B\hat \r$.
Therefore
\be\label{04}\nabla_x p(\hat \r)=\hat \r\nabla_x(B\hat \r)+\hat \r L(m_{\hat \r})r(x)^{-3}{\bf e}_r, \ x\in G,
\ee
where ${\bf e}_r=(\f{x_1}{r(x)}, \f{x_2}{r(x)}, 0)$. Moreover, it is proved in \cite{CF} that  the boundary $\pa G$ of $G$  is smooth enough
to apply the Gauss-Green formula (cf. \cite{evans}) on G. Applying the Gauss-Green formula on G and noting that $\hat \r|_{\pa G}=0$, we obtain,
\be\label{05}
\int_G x\cdot \nabla_x p(\hat \r)dx=-3\int_G p(\hat \r)dx=- 3\int p(\hat \r)dx.\ee
By an argument in \cite {Ta} (used also in \cite{LY}), we obtain
\be\label{jjyy} \int_G x\cdot \hat \r\nabla_x B\hat \r dx=-\f{1}{2}\int_G\hat \r B\hat \r dx=-\f{1}{2}\int\hat \r B\hat \r dx.
\ee
(In fact, this can be verified as follows. Let
$$I=\int_G x\cdot \hat \r\nabla_x B\hat \r dx=-\int_G \hat \r(x)\int_{G} \f{\r(y)(x-y)\cdot x}{|x-y|^3}dydx.$$
Then
\begin{align}
I&=-\int_G \hat \r(x)\int_{G} \f{\hat \r(y)(x-y)\cdot (x-y)}{|x-y|^3}dydx-\int_G \hat \r(x)\int_{G} \f{\r(y)(x-y)\cdot y}{|x-y|^3}dydx\notag\\
&=-\int_G\hat \r(x)\int_G \f{\hat \r(y)(x-y)\cdot (x-y)}{|x-y|^3}dydx-I\notag\\
&=-\int_G\hat \r B\hat \r dx-I,\end{align} which is (\ref{jjyy}).)
Next, since $x\cdot {\bf e}_r=r(x)$, we have
\begin{align}\label{07}
&\int_G x\cdot \hat \r(x) L(m_{\hat \r}(r(x))r^{-3}(x){\bf e}_rdx\notag\\&=\int_G \hat \r(x) L(m_{\hat \r}(r(x))r^{-2}(x)dx\notag\\&=\int \hat \r(x) L(m_{\hat \r}(r(x))r^{-2}(x)dx.\end{align}
Therefore,  from (\ref{05})-(\ref{07}) we have
\be\label{08}
-3\int p(\hat \r)dx=-\f{1}{2}\int\hat \r B\hat \r dx+\int \hat \r(x) L(m_{\hat \r}(r(x))r^{-2}(x)dx,
\ee
so that
$$ F(\hat \r)=\f{4-3\g}{\g-1}\int p(\hat \r)dx-\f{1}{2}\int \hat \r(x) L(m_{\hat \r}(r(x))r^{-2}(x)dx.$$
Thus, if $\g>4/3$, $F(\hat \r)<0$ since $L(m)\ge 0$ for $0\le m\le M$. Since $\hat \r\in W_{M, S}\subset W_M$, then $inf_{\r\in W_M}F(\r)<0.$  This completes the proof of part (a). \\
The proof of part (b) follows from a scaling argument as in \cite{rein1}. Taking $b=(M/\bar M)^{1/3}$ and letting $\bar \r(x)=\r(bx)$ for any
$\r\in W_M$. It is easy to verify that $\bar \r \in W_{\bar M}$ and that the following identities hold,
\be\label{010}
\int \bar \r B\bar \r dx=b^{-5}\int  \r B \r dx,
\ee
\be\label{0101}
\int  A(\bar\r )dx=b^{-3}\int  A(\r) dx.
\ee
Moreover, for $r\ge 0$,
\begin{align}
m_{\b \r}(r)&=2\pi \int_0^r s\int_{-\infty}^{\infty} \b \r(s, z)dsdz\notag\\
&=2\pi \int_0^r s\int_{-\infty}^{\infty}  \r(bs, bz)dsdz\notag\\
&=\f{1}{b^3}2\pi \int_0^{br} s'\int_{-\infty}^{\infty}  \r(s', z')ds'dz'\notag\\
&=\f{1}{b^3}m_{ \r}(br).\end{align}
Since $L$ satisfies (\ref{L1}) and $b> 1$,  we have
\be\label{L2}
L(m_{\bar \r}(r))\ge \f{1}{b^4} L(m_\r(br)).\ee
Thus,
\begin{align}\label{012}
\int\f{\b\r(x) L(m_{\b \r}(r(x)))}{r(x)^2}dx&\ge \f{1}{b^4}\int_0^{+\infty}\f{2\pi r}{r^2}L(m_{\r}(br))\int_{-\infty}^{\infty}\r(br, bz)dzdr\notag\\
&=\f{1}{b^5}\int_0^{+\infty}\f{2\pi r'}{r'^2}L(m_{\r}(r'))\int_{-\infty}^{\infty}\r(r', z')dz'dr'\notag\\
&=\f{1}{b^5}\int\f{\r(x) L(m_{\b \r}(r(x)))}{r(x)^2}dx.\end{align}
Therefore, since $b\ge 1$,  it follows from (\ref{010})-(\ref{012}) that
\begin{align} F(\b\r)&\ge b^{-3}\int A(\r)dx-\f{b^{-5}}{2}\int \r B\r dx+\f{b^{-5}}{2}\int\f{\r(x) L(m_{\b \r}(r(x)))}{r(x)^2}dx\notag\\
&\ge b^{-5}\left(\int A(\r)dx-\f{1}{2}\int \r B\r dx+\f{1}{2}\int\f{\r(x) L(m_{\b \r}(r(x)))}{r(x)^2}dx\right)\notag\\
&=(\b M/M)^{5/3} F(\r).\end{align}
Since $\r\to\b\r$ is one-to-one between $W_M$ and $W_{\b M}$, this proves part (b).

\end{proof}

The following lemma gives the boundedness of a minimizing sequence of $F$ in $L^{\g}(\RR^3)$.
\begin{lem}\label{lem4.3} Suppose $\g>4/3$. Let $\{\r^i\}\subset W_M $ be a minimizing sequence of $F$. Then $\{\r^i\}$ is bounded in
$L^{\g}(\RR^3)$,  and moreover, the rotating kinetic energy $$\f{1}{2}\int \f{\rho^i(x)L(m_{\r^i}(r(x)))}{r(x)^2}$$ is also uniformly bounded. \end{lem}
\begin{proof}
By Lemma \ref{lem4.2}, we have
\be\la{us1} \int [(\r^i)^{\g}(x)+\f{\rho^i(x)L(m_{\r^i}(r(x)))}{r(x)^2}] dx\le C_1 F(\r^i)+C_2,\  i\ge 1. \ee The lemma follows from this and Part a) in Lemma \ref{lem4.4}. \end{proof}

\begin{lem}\label{lem4.5} Suppose $\g>4/3$. Let $\{\r^i\}\subset W_M $ be a minimizing sequence for $F$. Then there exist constants
$r_0>0$, $\delta_0>0$,  $i_0\in \mathbf{N}$ and $x^i\in \RR^3$ with $r(x^i)\le r_0$,  such that
\be\la{keynote}
\int_{B_1(x^i)}\r^i(x)dx\ge \delta_0, \ i\ge i_0.
\ee\end{lem}
\begin{proof} 
First, since $\lim_{i\to\infty}F(\r^i)\to f_M$ and $f_M<0$ (see part (a) of Lemma \ref{lem4.4}), for large $i$,
\be\label{y1}
-\f{f_M}{2}\le -F(\r^i)\le \f{1}{2}\int \r^iB\r^idx.\ee
For any  $i$, let
\be \delta_i=\sup_{x\in \RR^3}\int_{|y-x|<1}\rho^i(y)dy.\ee
Now \begin{align}\label{y2'}
&\int \r^iB\r^i(x)dx\notag\\
&=\int_{\RR^3}\r^i(x)\int_{\RR^3} \f{\r^i(y)}{|y-x|}dydx\notag\\
&=\int_{\RR^3}\r^i(x)\int_{|y-x|<1}\f{\r^i(y)}{|y-x|}dydx+\int_{\RR^3}\r^i(x) \int_{1<|y-x|<r}\f{\r^i(y)}{|y-x|}dydx+\int_{\RR^3}\r^i(x)\int_{|y-x|>r}\f{\r^i(y)}{|y-x|}dydx\notag\\
&=:B_1+B_2+B_3,\end{align}
and $B_3\le M^2r^{-1}$. The shell $1<|y-x|<r$ can be covered by at most $ Cr^3$ balls of radius 1, so $B_2\le C M \delta_ir^3$.
By using H${\rm \ddot{o}}$lder's inequality and (\ref{bf1}), we get
\begin{align}\label{new111}
B_1&=\int_{\RR^3}\r^i(x)\int_{\RR^3}\f{\chi_{\{y||y-x|<1\}}(y)\r^i(y)}{|y-x|}dydx\notag\\
&\le \|\r^i\|_{4/3}\|B\{\chi_{\{y||y-x|<1\}}(y)\r^i(y)\}\|_4\notag\\
&\le C \|\r^i\|_{4/3}\left(\|\chi_{\{y||y-x|<1\}}(y)\r^i(y)\|_1^b\|\r^i\|_{4/3}^{1-b}+\|\chi_{\{y||y-x|<1\}}(y)\r^i(y)\|_1^c\|\r^i\|_{4/3}^{1-c}\right)\notag\\
&\le C \|\r^i\|_{4/3}\left(\delta_i^b\|\r^i\|_{4/3}^{1-b}+\delta_i^c\|\r^i\|_{4/3}^{1-c}\right),\end{align}
where $\chi$ is the indicator function, $0<b<1$ and $0<c<1$. By lemma 2.6, we know that $\|\r^i\|_{\g}$ is bounded, so $\|\r^i\|_{4/3}$ is bounded 
if $\g\ge 4/3$ in view of (2.17) and the fact $\|\r^i\|_1=M$. 
 This gives $B_1\le C(\delta_i^b+\delta_i^c)$. It follows that we could choose $r$ so large that the above estimates give $\int \r^iB\r^i(x)dx<-f_M$ {\it if $\delta_i$ were small enough}. This would contradict
(\ref{y1}). So there exists $\delta_0>0$ such that $\delta_i\ge \delta_0$ for large $i$. Thus, as $i$ is large, there
exists $x^i\in \RR^3$ and $i_0\in \mathbb{N}$ such that  \be\la{keynote1}
\int_{B_1(x^i)}\r^i(x)dx\ge \delta_0, \ i\ge i_0.
\ee   We now prove that there exists $r_0>0$ independent of $i$ such that those $x^i$ must satisfy
$r(x^i)\le r_0$ for $i$ large. Namely, since $\r^i$ has mass at least $\delta_0$ in the unit ball centered at $x^i$, and is
axially symmetric, it has mass $\ge Cr(x^i)\delta_0$ in the torus obtained by revolving this ball around $x_3$-axis (or $z$-axis).Therefore $r(x^i)\le (C\delta_0)^{-1}M.$
\end{proof}
In order to prove Theorem \ref{aa}, we will need the following lemma which is proved in \cite{rein1}, and uses a concentration-compactness argument.
\begin{lem}\label{lem4.6} Suppose $\g>4/3$. Let $\{f^i\}$ be a bounded sequence in $L^{\gamma}(\RR^3)$ and suppose
$$f^i\rightharpoonup f^0~~ weakly~in~ L^{\gamma}(\RR^3).$$ Then\\
(a) For any $R>0$,
$$\nabla B(\chi_{B_R(0)}f^i)\to \nabla B(\chi_{B_R(0)}f^0) ~~strongly~in ~ L^2(\RR^3),$$ where $\chi$ is the indicator function.\\
(b) If in addition $\{f^i\}$ is bounded in $L^1(\RR^3)$, $f^0\in L^1(\RR^3)$, and for any $\epsilon>0$ there exist $R>0$ and $i_0\in \mathbf N$ such that
\be\la{y3}\int_{|x|>R}|f^i(x)|dx<\epsilon,\qquad i\ge i_0,\ee
then
 $$\nabla Bf^i\to \nabla Bf^0 ~strongly~in ~ L^2(\RR^3).$$
\end{lem}
Before giving the proof of Theorem \ref{aa}, we first outline the main steps.   In step 1, we first show (2.15) and (2.16). In step 2 we show that if $\tilde \r$  is a weak limit in $L^{\g}(\RR^3)$ of $\{T\r^i\}$, then $m_{\tilde \r}(r)$ is a continuous function of $r$ for all
$r\ge 0$. The third step is to prove that $F$ is lower semi-continuous with respect to the weak topology in $L^\g(\RR^3)$.
\vskip  0.2cm
\noindent {\it Proof of Theorem \ref{aa}}
\vskip 0.2cm
\noindent\underline{Step 1}.
We prove (2.16), and apply Lemma \ref{lem4.6} to prove (2.14).    We begin with a splitting as in \cite{rein1}.
For $\r\in W_M$, for any $0<R_1<R_2$, we have
\be\label{017}
\r=\r\chi_{|x|\le R_1}+\r\chi_{R_1<|x|\le R_2}+\r\chi_{|x|>R_2} =:\r_1+\r_2+\r_3,\ee where $\chi$ is the indicator function.
It is easy to verify that
\be\label{018}
\int A(\r)dx=\sum_{j=1}^3\int A(\r_j)dx,\ee
and
\be \int \r B\r dx=\sum_{j=1}^3\int \r_jB \r_jdx+I_{12}+I_{13}+I_{23},\ee
where
$$I_{ij}=\int_{\RR^3}\int_{\RR^3} |x-y|^{-1}\r_i(x)\r_j(y)dxdy,\qquad 1\le i<j\le 3.$$
Also,
\begin{align}\la{020} \int\f{\r(x)L(m_{\r}(r(x))}{r^2(x)}dx&=\sum_{j=1}^3\int\f{\r_j(x)L(m_{\r_j}(r(x))}{r^2(x)}dx\notag\\
&+\sum_{j=1}^3\int\f{\r_j(x)(L(m_{\r}(r(x))-L(m_{\r_j}(r(x))}{r^2(x)}dx.\end{align}
It follows from (\ref{017})-(\ref{020}) that
\begin{align}\label{021}
F(\r)&=\sum_{j=1}^3F(\r_j)-\f{1}{2}\sum_{1\le i<j\le 3}I_{ij}\notag\\
&+\f{1}{2}\sum_{j=1}^3\int\f{\r_j(x)(L(m_{\r}(r(x))-L(m_{\r_j}(r(x))}{r^2(x)}dx.
\end{align}
Since $\r\ge \r_j$ , we have $m_\r(r)\ge m_{\r_j}(r)$ for any $r\ge 0$ and $j=1, 2, 3$. By (\ref{L2'}),
\be\label{022}
F(\r)\ge \sum_{j=1}^3F(\r_j)-\f{1}{2}\sum_{1\le i<j\le 3}I_{ij}.\ee
Using (\ref{022}) and Lemma \ref{lem4.4}, by the same argument as in the proof of Theorem 3.1 in \cite{rein1}, we can show that
\be\label{rx1} f_M-F(\rho)\le C f_M M_1M_3+C(R_2^{-1}+||\r||_{\g}^{(q+1)/6}||\nabla B\r_2||_2),\ee
by choosing $R_2>2R_1$ in the splitting (\ref{017}), where $M_1=\int \r_1(x)dx=\int_{|x|\le R_1}\r(x)dx$,  $M_3=\int \r_3(x)dx=\int_{|x|> R_2}\r(x)dx$
and $q=1/(\g-1)$.
 Let $\{\r^i\}$ be a minimizing sequence of $F$ in $W_M$. By Lemma \ref{lem4.5}, we know that there exists  $i_0 \in \mathbf{N}$ and $\delta_0>0$ independent of $i$ such that
 \be\la{y5}
\int_{B_1(x^i)}\r^i(x)dx\ge \delta_0, \qquad i\ge i_0 \ee
for some $x^i\in \RR^3$ with $r(x^i)\le r_0$ for some constant $r_0>0$ independent of $i$.
Let $a_i=z(x^i)$ and $R_0=r_0+1$, then (\ref{y5}) implies
\be\label{y4}
\int_{a_i{\bf e_3}+B_{R_0(0)}}\r^i(x)dx\ge \delta_0,  \qquad  if~i\ge i_0,\ee
where ${\bf e_3}=(0, 0,1)$.
Having proved (\ref{y4}), we can  follow the argument in the proof of Theorem 3.1 in \cite{rein1} to verify (\ref{y3}) for
$$f^i(x)=T\r^i(x)=:\r^i(\cdot+a_i{\bf e_3})$$ by using (\ref{022}) and (\ref{y4}) and choosing
suitable $R_1$ and $R_2$ in the splitting (\ref{017}). We sketch this as follows. The sequence $T\r^i=:\r^i(\cdot+a_i{\bf e_3})$, $i\ge i_0$, is a minimizing sequence of $F$ in $W_M$ (see Remark 2 after Theorem \ref{aa}). We rewrite (\ref{y4}) as
\be\label{y4'}
\int_{B_R(0)}T\r^i(x)dx\ge \delta_0, \  i\ge i_0.\ee
 Applying (\ref{rx1}) with $T\r^i$ replacing $\r$,  and noticing that $\{T\r^i\}$ is bounded in $L^\g(\RR^3)$ (see Lemma 2.6),  we obtain, if $R_2>2R_1$,
\be\la{rx3} -C f_M M^i_{1}M^i_{3}\le C(R_2^{-1}+||\nabla BT\r^i_{2}||_2)+F(T\r^i)-f_M,\ee
where $M^i_{1}=\int T\r^i_1(x)dx=\int_{|x|<R_1}T\r^i(x)dx,$, $M^i_{3}=\int T\r^i_3(x)dx=\int_{|x|>R_2}T\r^i(x)dx$ and $T\r^i_{2}=\chi_{R_1<|x|\le R_2}T\r^i.$ Since $\{T\r^i\}$ is bounded in $L^{\g}(\RR^3)$, there exists a subsequence, still labeled by $\{T\r^i\}$, and a function $\tilde \r\in W_M$
such that $$T\r^i\rightharpoonup \tilde \r{~\rm weakly~ in~} L^{\g}(\RR^3).$$  This proves (2.15). By (\ref{y4'}), we know that $M^i_{1}$ in (\ref{rx3}) satisfies $M^i_{1}\ge \delta_0$
for $i\ge i_0$ by choosing $R_1\ge R_0$ where $R_0$ is the constant in (\ref{y4'}). Therefore, by (\ref{rx3}) and the fact that $f_M<0$ (cf. Part (a) in Lemma 2.5) , we have
\be\la{rx4} -C f_M \delta_0 M^i_{3}\le CR_2^{-1}+C||\nabla B\tilde \r_{2}||_2+C||\nabla BT\r^i_{2}-\nabla B\tilde \r_{2}||_2)+F(T\r^i)-f_M,\ee
where  $\tilde \r_{2}=\chi_{|x|>R_2}\tilde \r$.  Given any $\epsilon>0$, by the same
argument as \cite{rein1}, we can increase $R_1>R_0$ such that the second term on the right hand side of (\ref{rx4}) is small, say less than $\epsilon/4$.
Next choose $R_2>2R_1$ such that the first term is small. Now that $R_1$ and $R_2$ are fixed, the third term on the right hand side of (\ref{rx4}) converges to zero by Lemma \ref{lem4.6}(a).  Since $\{T\r^i\}$ is a minimizing sequence of $F$ in $W_M$, we can make $F(T\r^i)-f_M$ small by taking $i$ large.
Therefore, for $i$ sufficiently large, we can make
\be\label{rx5}  M^i_{3}=:\int_{|x|>R_2}T\r^i(x)dx<\epsilon.\ee
This verifies (\ref{y3}) in Lemma 2.8 for $f^i=T\r^i$. By weak convergence we have that for any $\epsilon>0$ there exists $R>0$ such that
$$M-\epsilon\le \int_{B_R(0)}\tilde \r(x)dx\le M,$$
which implies $\tilde \r\in L^1(\RR^3)$ with $\int \tilde \r dx=M$. Therefore,  by Lemma \ref{lem4.6}(b),we have
\be\label{rx6} ||\nabla BT\r^i-\nabla B\tilde \r||_2\to 0, \qquad i\to +\infty.\ee
This proves (2.16).    (2.14) in Theorem \ref{aa} follows from (\ref{rx5}) by taking $R=R_2$.
\vskip 0.2cm
\noindent\underline{Step 2}. Let  $\tilde \r$ be a weak limit of a subsequence of $\{T\r^i\}$ in $L^{\gamma}(\RR^3)$ (we still label the subsequence by $\{T\r^i\}$). We claim that the mass function
 \be\label{pr1} m_{\tilde \r}(r)=:\int_{\sqrt {x_1^2+x_2^2}\le r} \tilde \r(x) dx{\rm~~~ is~ continuous~ for~} r\ge 0.\ee
This is proved as follows.
By the lower semicontinuity of norms (cf. \cite{lieb} p.51) and Lemma 2.6,
we have
\be\label{pr2}||\tilde \r||_{\g}\le \lim_{i\to \infty}\inf ||T\r^i||_{\g}=\lim_{i\to \infty}\inf ||\r^i||_{\g}\le C,
\ee
for some positive constant $C$. For any $\epsilon>0$,  by the weak convergence and (\ref{2.15}) which we have already proved,  there exists $R>0$ such that
\be\label{rx9}
\int_{|x|> R} T\r^i(x)dx<\epsilon, \qquad i\in \mathbb{N},\ee and
\be\label{rx8}
\int_{|x|> R} \tilde \r(x)dx=\lim_{i\to \infty}\int_{|x|> R} T\r^i(x)dx\le \epsilon.\ee
For any $r\ge 0$ and $r_1\ge r$,
\begin{align}\label{pr3}
&0\le m_{\tilde \r}(r_1)-m_{\tilde \r}(r)\notag\\&=\int_{r\le \sqrt{x_1^2+x_2^2}\le r_1}\tilde \r (x)dx\notag\\
&\int_{r\le \sqrt{x_1^2+x_2^2}\le r_1, |x_3|> R}\tilde \r (x)dx+\int_{r\le \sqrt{x_1^2+x_2^2}\le r_1, |x_3|\le R}\tilde \r (x)dx.
\end{align}
Since $\{x=(x_1, x_2, x_3)\in\RR^3: r\le \sqrt{x_1^2+x_2^2}\le r_1, |x_3|> R\}\subset \{x=(x_1, x_2, x_3)\in\RR^3:  |x|> R\}$, by (\ref{rx8}), we have
\be\label{pr4} \int_{r\le \sqrt{x_1^2+x_2^2}\le r_1, |x_3|> R}\tilde \r (x)dx<\epsilon.
\ee
By (\ref{pr2}) and H${\rm\ddot{o}}$lder's inequality,
\begin{align}\label{pr5}
&\int_{r\le \sqrt{x_1^2+x_2^2}\le r_1, |x_3|\le R}\tilde \r (x)dx\notag\\
&\le ||\tilde \r||_{\g}\left(meas\{x=(x_1, x_2, x_3)\in \RR^3:  r\le \sqrt{x_1^2+x_2^2}\le r_1, |x_3|\le R\}\right)^{1/\g'}\notag\\
&\le C [2\pi R (r_1+r)(r_1-r)]^{1/\g'},
\end{align}
where $meas$ denotes the Lebsgue measure and $\g'=(\g-1)/\g.$ Now, if we take $\delta=\min\{\f{(\epsilon/C)^{\g'}}{2\pi R(2r+1)}, 1\}$, then by (\ref{pr5}),
we obtain
\be\label{pr6}\int_{r\le \sqrt{x_1^2+x_2^2}\le r_1, |x_3|\le R}\tilde \r (x)dx<\epsilon,\ee
whenever $0\le r_1-r<\delta.$
It follows from (\ref{pr3}), (\ref{pr4}) and ({\ref{pr5}), we have
\be\label{pr61}
| m_{\tilde \r}(r_1)-m_{\tilde \r}(r)|<2\epsilon,
\ee
whenever $0\le r_1-r<\delta.$ This proves that $m_{\tilde \r}(r)$ is continuous from the right for any $r\ge  0$. By the same method, we can show that $m_{\tilde \r}(r)$ is continuous from the left for any $r> 0$ . Since $m_{\tilde \r}(0)=0$,  this proves (\ref{pr1}).
\vskip 0.2cm
\noindent\underline{Step 3}. Let  $\{\r^i\}$ be a minimizing sequence of the energy functional  $F$, and let $\tilde \r$ be a weak limit  of $\{T\r^i\}$ in $L^{\gamma}(\RR^3)$.  We will prove that $\tilde \r$ is a minimizer of $F$ in $W_M$; that is
\be\label{2.80}
F(\tilde \r)\le \lim\inf_{i\to \infty} F(T\r^i).\ee
First, by (\ref{pr2}), we have
\be\label{rx7} \int A(\tilde \r)dx\le \lim\inf_{i\to \infty} \int A(T\r^i)dx.\ee
 We fix a positive number $\delta$ and show that
\be\label{rx10}\lim_{i\to \infty}\int_{r(x)\ge \delta}\f{T\r^i(x)L(m_{T\r^i}(r(x))-\tilde \r(x)L(m_{\tilde \r}(r(x))}{r^2(x)}dx=0.\ee
To see this, we write
\begin{align}\label{rx11}
&\int_{r(x)\ge \delta}\f{(T\r^i(x)L(m_{T\r^i}(r(x))-\tilde \r(x)L(m_{\tilde \r}(r(x))}{r^2(x)}dx\notag\\
&=\int_{r(x)\ge \delta}\f{(T\r^i(x)-\tilde \r(x))L(m_{\tilde \r}(r(x))}{r^2(x)}dx\notag\\&+\int_{r(x)\ge \delta}\f{T\r^i(x)(L(m_{T\r^i}(r(x))-L(m_{\tilde \r}(r(x)))}{r^2(x)}dx.
\end{align} For any $R>0$, we have
\begin{align}\label{rx12}
&\int_{r(x)\ge \delta}\f{(T\r^i(x)-\tilde \r(x))L(m_{\tilde \r}(r(x))}{r^2(x)}dx\notag\\
&=\int_{r(x)\ge \delta, |x|\le R}\f{(T\r^i(x)-\tilde \r(x))L(m_{\tilde \r}(r(x))}{r^2(x)}dx\notag\\&+\int_{r(x)\ge \delta, |x|\ge  R}\f{(T\r^i(x)-\tilde \r(x))L(m_{\tilde \r}(r(x))}{r^2(x)}dx.\end{align}
\vskip 0.1cm
\noindent In view of  (\ref{rx9}) and (\ref{rx8}), for any $\epsilon>0$,  we can choose $R$ such that
\vskip 0.1cm
\be\label{rx13}|\int_{r(x)\ge \delta, |x|\ge R}\f{(T\r^i(x)-\tilde \r(x))L(m_{\tilde \r}(r(x))}{r^2(x)}dx|
\le \f{2L(M)\epsilon}{\delta^2}.\ee
\vskip 0.1cm
\noindent By the weak convergence of $\{T\r^i\}$ in  $L^{\g}(\RR^3)$ and the fact that $L$ is defined on a bounded range, $L(m_{\tilde \r}(r(x))\chi_{\{r(x)\ge \delta, |x|\le R\}}(x){r^{-2}(x)}\in
L^{\g'}(\RR^3)$, where as before $\chi$ is the indicator function, and  $\g'=\f{\g}{\g-1}$ (satisfying $1/\g+1/\g'=1$). We have
\begin{align}\label{rx14}
&\lim_{i\to \infty}\int_{r(x)\ge \delta, |x|\le R}\f{(T\r^i(x)-\tilde \r(x))L(m_{\tilde \r}(r(x))}{r^2(x)}dx\notag\\
&=\lim_{i\to \infty}\int (T\r^i(x)-\tilde \r(x))L(m_{\tilde \r}(r(x))\chi_{\{r(x)\ge \delta, |x|\le R\}}(x){r^{-2}(x)}dx\notag\\
&=0, \end{align}
because $T\r^i$ converges weakly to $\tilde \r$.
Since $\epsilon$ is arbitrary, (\ref{rx13}) and (\ref{rx14}) imply
\be\label{rx15}
\lim_{i\to \infty}\int_{r(x)\ge \delta}\f{(T\r^i(x)-\tilde \r(x))L(m_{\tilde \r}(r(x))}{r^2(x)}dx=0.\ee
We handle the second term in (\ref{rx11}) as follows. By weak convergence, we know that $m_{T\r^i}(r)$ converges to $m_{\tilde \r}(r)$ pointwise for $r\ge 0$.  Since $m_{T\r^i}(r)$ and $m_{\tilde \r}(r)$ are non-decreasing functions of $r$ for $r\ge 0$ and $m_{\tilde \r}(r)$ is continuous on $[0, +\infty)$ (see (\ref{pr1})), by a variation on Dini's theorem  (\cite{Rudin}, p.167)$^*$, \begin{figure}[b]\rule[-2.5truemm]{5cm}{0.1 truemm}
{\footnotesize \\
$^*$ We thank Dmitry Khanvinson for pointing out this to us. }
\end{figure}  we know that $m_{T\r^i}(r)$ converges to $m_{\tilde\r}(r)$ uniformly on the interval $ [0, R]$ for any $R>0$.   Since $L\in C^1[0, M]$,  it follows that $L(m_{T\r^i}(r))$ converges to $L(m_{\tilde \r}(r))$ uniformly on any interval $[0, R]$. For any $\epsilon>0$, we can fix $R>0$ such that (\ref{rx9}) and (\ref{rx8}) hold.  Since $L(m_{T\r^i}(r))$ converges uniformy to $L(m_{\tilde \r}(r))$ on any interval $[0, R]$, we have
\be\label{pr9}
\lim_{i \to \infty}||L(m_{T\r^i}(\cdot))-L(m_{\tilde \r}(\cdot))||_{L^{\infty}[0, R]}=0.
\ee
Let
\be\label{rx17}
A_{\delta}=\{x\in\RR^3, r(x)\ge \delta\},
\ee
then we have, using (\ref{rx9}) and (\ref{rx8})  that
\begin{align}\label{rx18}
&|\int_{r(x)\ge \delta}\f{T\r^i(x)(L(m_{T\r^i}(r(x))-L(m_{\tilde \r}(r(x)))}{r^2(x)}dx|\notag\\
&\le |\int_{A_\delta\cap B_{R}(0)}\f{T\r^i(x)(L(m_{T\r^i}(r(x))-L(m_{\tilde \r}(r(x)))}{r^2(x)}dx|\notag\\
&+|\int_{A_\delta-B_{R}(0)}\f{T\r^i(x)(L(m_{T\r^i}(r(x))-L(m_{\tilde \r}(r(x)))}{r^2(x)}dx|\notag\\
&\le  ||L(m_{T\r^i}(\cdot))-L(m_{\tilde \r}(\cdot))||_{L^{\infty}[0, R]}\delta^{-2}M
+2\delta^{-2}L(M) \epsilon.
\end{align}
Since $\epsilon$ is arbitrary, it follows from (\ref{pr9}) and (\ref{rx18}) that
\be\label{rx19}\lim_{i\to \infty}\int_{r(x)\ge \delta}\f{T\r^i(x)(L(m_{T\r^i}(r(x))-L(m_{\tilde \r}(r(x)))}{r^2(x)}dx=0.
\ee
This, together with (\ref{rx11})and (\ref{rx15}), implies (\ref{rx10}).
Next, we show that
\be\label{rx20}
\lim_{i\to \infty}\inf\int\f{T\r^i(x)L(m_{T\r^i}(r(x))-\tilde \r(x)L(m_{\tilde \r}(r(x))}{r^2(x)}dx\ge 0,
\ee
by using (\ref{rx10}) and the monotone convergence theorem for integrals. In fact we have
\begin{align}\label{rx21}
&\int\f{T\r^i(x)L(m_{T\r^i}(r(x))-\tilde \r(x)L(m_{\tilde \r}(r(x)))}{r^2(x)}dx\notag\\
&=\int\f{T\r^i(x)L(m_{T\r^i}(r(x)))(1-\chi_{A_\delta})}{r^2(x)}dx\notag\\
&+\int\f{[T\r^i(x)L(m_{T\r^i}(r(x)))-\tilde \r(x)L(m_{\tilde \r}(r(x)))]\chi_{A_\delta}}{r^2(x)}dx\notag\\
&+\int\f{\tilde \r(x)L(m_{\tilde \r}(r(x)))(\chi_{A_\delta}-1)}{r^2(x)}dx,
\end{align}
where $\chi$ is the indicator function, and $A_\delta$ is the set defined in (\ref{rx17}).
For any $ i\ge 1$,
\be\label{rx21'} \int\f{T\r^i(x)L(m_{T\r^i}(r(x)))(1-\chi_{A_\delta})}{r^2(x)}dx\ge 0.
\ee
We fix $\delta$, and by (\ref{rx10}), we know that
the second term  on the right hand side of (\ref{rx21}) approaches zero as $i\to \infty$.
Therefore, in view of (\ref{rx21'}),
\begin{align}\label{rx22}
&\lim\inf_{i\to \infty}\int\f{T\r^i(x)L(m_{T\r^i}(r(x)))-\tilde \r(x)L(m_{\tilde \r}(r(x)))}{r^2(x)}dx\notag\\
&\ge \int\f{\tilde \r(x)L(m_{\tilde \r}(r(x)))(\chi_{A_\delta}-1)}{r^2(x)}dx.
\end{align}
By the monotone convergence theorem of integrals, we have
\be\label{rx23}
\lim_{\delta\to 0}|\int\f{\tilde \r(x)L(m_{\tilde \r}(r(x)))(\chi_{A_\delta}-1)}{r^2(x)}dx|=0.
\ee
Letting $\delta\to 0$ in (\ref{rx22}), gives (\ref{rx20}). By (\ref{rx6}), (2.71}) and (\ref{rx20}), we obtain
\be\label{2.801}
F(\tilde \r)\le \lim\inf_{i\to \infty} F(T\r^i).\ee
Since $T\r^i$ is a minimizing sequence,  $\tilde \r$ is a minimizer of $F$ in $W_M$. This completes the proof of Theorem \ref{aa}.

\section{Nonlinear Stability of Rotating Star Solutions}.

We consider the Cauchy problem for
(\ref{1.1}) with the initial data
\be\label{initial}\rho(x,0)=\rho_0(x),\
{\bf v}(x,0)={\bf v}_0(x).\ee
We begin by giving the definition of a weak solution.\\

\noindent {\bf Definition:}  Let $\r {\bf v}={\bf m}$. The triple $(\rho, {\bf m}, \Phi)(x, t)$ ($x\in\RR^3, t\in[0, T])$ $(T>0)$  and  $\Phi$ given by (\ref{phi},  with $\r\ge 0,$ ${\bf m}$, ${\bf m}\otimes{\bf m}/\r$ and $ \r\nabla\Phi$ being in  $L^1_{loc}( \RR^3\times [0, T])$, is called a  {\it weak solution} of the Cauchy problem  (\ref{1.1}) and (\ref{initial}) on $ \RR^3\times [0, T]$  if for any Lipschitz continuous test  functions $\psi$ and ${\bf \Psi}=(\psi_1, \psi_2, \psi_3)$ with compact supports in $\RR^3\times [0, T]$,\\
  \be \int_0^T\int \left(\rho\psi_t+{\bf m}\cdot \nabla\psi\right)dxdt+\int\rho_0(x)\psi(x,0)dx=0,
 \ee
 and
 \be\label{3.3} \int_0^T\int \left({\bf m}\cdot{\bf \Psi}_t+\f{{\bf m}\otimes{\bf m}}{\rho}\cdot \nabla{\bf \Psi}\right)dxdt+\int{\bf m}_0(x){\bf \Psi}(x,0)dx=\int_0^T
 \int\rho\nabla \Phi{\bf \Psi} dxdt,
 \ee both hold.

\vskip 0.2cm

For any weak solution, it is easy to verify that the total mass is conserved by using a generalized divergence theorem for
 $L^{r}$ functions ($r\ge 1$) (cf. \cite{chenfrid}),
 \be\label{5.1}
 \int\r(x, t)dx=\int \r(x, 0)dx,\qquad t\ge 0.\ee
The {\it total energy} of system (\ref{1.1}) at time $t$ is
\begin{equation}\label{energy}
E(t)=E(\r(t), {\bf v}(t))=\int\left(A(\r)+\frac{1}{2}\r|{\bf v}|^2\right)(x, t)dx-\frac{1}{8\pi}\int|\nabla \Phi|^2(x, t)dx,\end{equation}
where as before,
\be\label{A}A(\r)=\f{p(\r)}{\g-1}.\ee
Note that the energy $E(t)$   has both a positive
and a negative part. This makes the stability analysis highly nontrivial, as noted in \cite{Rein}.
For a solution of (\ref{1.1}) without shock waves,  the total energy is conserved,
i.e., $E(t)=E(0)$ ($t\ge0$)(cf. \cite{Ta}). For  solutions with shock waves, the energy should be non-increasing in time,
so that for all $t\ge 0$,
\be\label{denergy} E(t)\le E(0),\ee
due to the entropy conditions, which are motivated by the second law of thermodynamics (cf. \cite{lax} and \cite{smoller}). This will be proved in Theorem
5.1, below.

 We consider axi-symmetric initial data, which takes the form
 \begin{align}\la{5.2'}
  &\r_0(x)=\r(r, z),\notag\\
  & {\bf v}_0(x)=v^r_0(r, z){\bf e}_r+v^{\theta}_0(r, z){\bf e}_{\theta}+v^3_0(\r, z){\bf e}_3.
 \end{align}
  Here   $r=\sqrt {x_1^2+x_2^2},\ z=x_3$,  $x=(x_1, x_2,  x_3)\in \RR^3$ (as before), and
  \be {\bf e}_r=(x_1/r, x_2/r,  0)^\mathrm{T},\ {\bf e}_{\theta}=(-x_2/r,  x_1/r,\ 0)^\mathrm{T},\ {\bf e}_3=(0, 0, 1)^\mathrm{T}.\ee
  We seek axi-symmetric  solutions of the form
  \begin{align}\la{5.3'}
  &\r(x, t)=\r(r, z, t),\notag\\
  & {\bf v}(x, t)=v^r(r, z, t){\bf e}_r+v^{\theta}(r, z, t){\bf e}_{\theta}+v^3(r, z, t){\bf e}_3,\\
  &\Phi(x, t)=\Phi(r, z, t)=-B\r(r, z, t),
  \end{align}
 We call a vector field ${\bf u}(x, t)=(u_1, u_2, u_3)(x)$ ($x\in \RR^3$ ) axi-symmetric if it can be written
in the form
$${\bf u}(x)=u^r(r, z){\bf e}_r+u^{\theta}(r, z){\bf e}_{\theta}+u^3(\r, z){\bf e}_3.$$
For the velocity field ${\bf v}=(v_1, v_2, v_3)(x, t)$, we define the angular momentum  $j(x,t)$ about the $x_3$-axis  at $(x, t)$ ,  $t\ge 0$, by
\be\la{5.3}
j(x, t)=x_1v_2-x_2v_1.\ee
For an axi-symmetric velocity field
\be\la{asv}
{\bf v}(x, t)=v^r(r, z, t){\bf e}_r+v^{\theta}(r, z, t){\bf e}_{\theta}+v^3(\r, z, t){\bf e}_3,\ee
\be\la{comp}
v_1=\f{x_1}{r}v^r-\f{x_2}{r}v^{\theta},\ v_2=\f{x_2}{r}v^r+\f{x_1}{r}v^{\theta}, v_3=v^3,\ee
so that
\be\la{j}j(x, t)= r v^{\theta}(r, z,  t). \ee
In view of ( {\ref{asv}) and (\ref{j}), we have
\be\label{V}
|{\bf v}|^2=|v^r|^2+\f{j^2}{r^2}+|v^3|^2.\ee
Therefore, the total energy at time $t$ can be written as
\begin{align}\label{en}
E(\r(t), {\bf v}(t))
&=\int A(\r)(x, t)dx+\frac{1}{2}\int \f{\r j^2(x,t)}{r^2(x)}dx\notag\\
&-\frac{1}{8\pi}\int|\nabla B\r|^2(x, t)dx+\frac{1}{2}\int \r(|v^r|^2+|v^3|^2)(x, t)dx.\end{align}

There are two important conserved quantities for the Euler-Poisson equations (\ref{1.1}); namely  the total mass and the angular momentum. In order to
describe these, we define $D_t$,  the non-vacuum region at time $t\ge 0$ of the solution by
\be\label{nonvacuum}
D_t=\{x\in \RR^3: \r(x, t)>0\}.
\ee
We will make the following physically reasonable assumptions A1)-A4)  on weak solutions of the Cauchy problem (\ref{1.1}) and (\ref{initial}):
\vskip 0.2cm
 A1) For any $t\ge0$, there exists a measurable subset $G_t\subset D_t$ with $meas(D_t-G_t)=0$ ($meas$ denotes the Lebsegue measure)  such that,
for any $x\in G_t$, there exists a unique (backwards) particle path $\xi(\tau, x, t)$ for $0\le \tau\le t$ satisfying

\be\label{particlepath}
\pa_{\t}\xi(\t, x, t)={\bf v}(\xi(\t, x, t), \t),\ \xi(t, x, t)=x.\ee

\vskip 0.2cm

For  $x\in G_t$, we write $$\xi(0, x, t)=\xi_{-t}(x).$$ Also, for $x\in \RR^3$ and $t\ge 0$,  we denote the total mass at time $t$
in the cylinder $\{y\in \RR^3: r(y)\le r(x)\}$ by $m_{\r(t)}(r(x))$, i.e.,
\be\label{mass}
m_{\r(t)}(r(x))=\int_{r(y)\le r(x)}\r(y, t)dy.\ee
For  axi-symmetric motion, we assume

\vskip 0.2cm

A2)
\be\label{mass1}
m_{\r(t)}(r(x))=m_{\r_0}(r(\xi_{-t}(x))), \qquad {\rm for~}  x\in G_t, t\ge 0.\ee
Moreover, the angular momentum  is conserved along the particle path:

\vskip 0.2cm

A3) \be\label{angular1}j (x, t)=j( \xi_{-t}(x), 0), \qquad {\rm for~}  x\in G_t, t\ge 0.\ee
(Both (3.21) and (3.22) are shown in \cite{Ta} if the solution has some  regularity.)

\vskip 0.2cm

\noindent Finally, for $L=j^2$,  we need a technical assumption; namely, \\
A4) \be\label{extra1}
\lim_{r\to 0+}\frac{L(m_{\r(t)}(r)+m_{\tilde \r}(r))m_{\sigma(t)}(r)}{r^2}=0,
\ee
for $t\ge 0$, where $\sigma(t)=\r(t)-\tilde \r. $
\begin{rem} (\ref{extra1})  can be understood as follows. For any $\r\in W_M$, we have $\lim_{r\to 0+} m_{\r}(r)=0. $ Therefore $\lim_{r\to 0+}L(m_{\r(t)}(r)+m_{\tilde \r}(r))=L(0)=0,$
so if we define $$\hat \r(s, t)-\hat \tilde \r(s)=\int_{-\infty}^{+\infty} (\r(s,z, t)- \tilde \r(s,z))dz, $$
then if
\be\label{good5} \f{m_{\sigma(t)}(r)}{r^2}=\f{\int_0^r(2\pi s (\hat \r(s, t)-\hat {\tilde \r}(s))ds}{r^2}\in L^{\infty}(0, \delta) \  for\  some \  \delta>0,
\ee
  (\ref{extra1})will hold.
If $\hat \r(\cdot, t)-\hat \tilde \r(\cdot)\in L^{\infty}(0, \delta)$, then  (\ref{good5}) holds. This can be assured by assuming that
$\r(r, z, t)-\tilde \r(r, z)\in L^{\infty}((0, \delta)\times \RR\times \RR^+)$ and decays  fast enough in the $z$ direction. For example,
when $\r(x, t)-\tilde \r(x)$ has  compact support in $\RR^3$ and $\r(\cdot, t)-\tilde \r(\cdot)\in L^{\infty}(\RR^3)$, then (\ref{extra1}) holds.
\end{rem}

\vskip 0.2cm

Now we make some assumptions on the initial data; namely,  we assume that the initial data is such that the initial total mass and
angular momentum are the same as those of the rotating star solution (those two quantities are conserved quantities). Therefore,
we require
\vskip 0.2 cm
I$_1$) \be\label{initial mass}
\int \r_0(x)dx=\int \tilde \r(x)dx=M. \ee
Moreover we assume
\vskip 0.2cm

I$_2$) For the initial angular momentum $j (x, 0)=rv_0^{\theta}(r, z)=: j_0(r, z)$ ($r=\sqrt {x_1^2+x_2^2}$, $z=x_3$ for $x=(x_1, x_2, x_3)$,
we assume
$j(x, 0)$ only depends on the total mass in the cylinder $\{y\in\RR^3, r(y)\le r(x)\}$, i.e. ,
\be\label{ia}
j(x, 0)=j_0\left(m_{\r_0}(r(x))\right).\ee
Finally, we assume that the initial profile of the angular momentum per unit mass is the same as that of the rotating star solution, i. e.,
\vskip 0.2cm
 I$_3$) \be\label{ia1}
j_0^2(m)=L(m), \qquad 0\le m\le M,\ee
where $L(m)$ is the profile of the square of the angular momentum of the rotating star defined in Section 2.
((\ref{ia}) implies that we require that $v_0^{\theta}(r, z)$ only depends on $r$.)\\
In order to state our stability result, we need some notation.
Let $\lambda$ be the number in Theorem 2.2, i.e.,
\be\label{lam}
\begin{cases}
& A'(\tilde \r(x))+\int_{r(x)}^{\infty}L(m_{\tilde \r}(s))s^{-3}ds-B\tilde \r(x)=\lambda, \  x\in \Gamma,\\
&\int_{r(x)}^{\infty}L(m_{\tilde \r})(s))s^{-3}ds-B\tilde \r(x)\ge \lambda, \qquad x\in \RR^3-\Gamma,\end{cases}\ee
with $A$ defined in (2.3), and $\Gamma$ defined in  (2.10).\\

For $\r\in L^1\cap L^{\g}$, we define, \be d(\r, \tilde \r)=\int
[A(\rho)-A(\tilde \r)]
+(\r-\tilde \r)\int_{r(x)}^{\infty}\{\f{L(m_{\tilde \r}(s))}{s^3}ds-\lambda-B\tilde \r\}dx.
\ee
\begin{rem} For $x\in \Gamma$, in view of (\ref{A}) and (3.28), we have,
\begin{align}
&(A(\rho)-A(\tilde \r))(x)
+(\int_{r(x)}^{\infty}\f{L(m_{\tilde \r}(s))}{s^3}ds-\lambda-B\tilde \r(x))(\r-\tilde \r)\notag\\
&= (A(\r)-A(\tilde \r)-A'(\tilde \r)(\r-\tilde \r))(x)\notag\\
&= \f{p(\rho)-p(\tilde \r)-p'(\tilde \r)(\r-\tilde \r)}{\g-1}(x)\ge 0.\end{align}
Thus, for $\r\in W_M$,
\be d(\r, \tilde \r)\ge 0. \ee
Moreover, $d(\r, \tilde \r)= 0$ if and only if $\r=\tilde \r$, and if $\g\le 2$,
\be d(\r, \tilde \r)\ge C||\r-\tilde \r||_2^2, \qquad \r\in W_M.\ee\end{rem}
We also define
\begin{align}\label{d1}d_1(\r, \tilde \r)
&=\f{1}{2}\int\f{\r(x) L(m_{\r}(r(x))-\tilde \r(x) L(m_{\tilde \r}(r(x))}{r^2(x)}dx\notag\\
&-\int \int_{r(x)}^{\infty}s^{-3}L(m_{\tilde \r}(s))ds(\r(x)-\tilde \r(x))dx,
\end{align}
for $\r\in W_M$. We shall show later that $d_1\ge 0$.
Our main stability result in this paper is the following global-in-time stability theorem.
\begin{thm}\label{th5.1} Let $\tilde \r$ be a minimizer of the functional $F$ in $W_M$, and assume that it  is unique up to a vertical shift.  Suppose $\g>4/3$ and  the above assumptions A1)-A4) and I$_1$)- I$_3$) hold. Moreover,  assume that the angular momentum  of the rotating star solution $\tilde \r$ satisfies  (\ref{L}), (\ref{L1}) and (2.13).  Let $(\r, {\bf v}, \Phi)(x, t)$ be an axi-symmetric weak solution of the Cauchy problem (\ref{1.1}), (\ref{initial}) with $\r(\cdot, t)\in L^1\cap L^{\g}$, $\r|{\bf v}|^2(\cdot, t)\in L^1$ and $\nabla\Phi(\cdot, t)=-\nabla B\r(\cdot, t)\in L^2$.  If the total energy $E(t)$ (c.f.  (3.5)) is non-increasing with respect to $t$,
then for every $\epsilon>0$, there exists a number $\delta>0$ such that if
\begin{align} &d(\r_0, \tilde \r)+\f{1}{8\pi}||\nabla B\r_0-\nabla B\tilde \r||_2^2+ |d_1(\r_0, \tilde \r)|\notag\\
&+\f{1}{2}\int \r_0(x)(|v^r_0|^2+|v^3_0|^2)(x)dx
<\delta,\end{align}
then  there is a vertical shift $a{\bf e_3}$ ($a\in \RR$, ${\bf e_3}=(0, 0, 1)$) such that, for every $t>0$
\begin{align} &d(\r(t), T^a\tilde \r)+\f{1}{8\pi}||\nabla B\r(t)-BT^a\tilde \r||_2^2+|d_1(\r(t), T^a\tilde \r)|\notag\\
&+\f{1}{2}\int \r(x, t)(|v^r(x, t)|^2+|v^3(x, t)|^2)dx
<\epsilon,
\end{align}
where $T^a\tilde \r(x)=:\tilde \r(x+a{\bf e_3}).$
\end{thm}
\begin{rem} The vertical shift $a{\bf e_3}$ appearing in the theorem is analogous to a similar phenomenon which appears in the study of stability of viscous traveling waves in conservation laws, whereby convergence is to a ``shift`` of the original traveling wave.
\end{rem}
\begin{rem}Without the uniqueness assumption for the minimizer of $F$  in $W_M$, we can have the following type of stability result, as observed
in \cite{Rein} for the non-rotating star solutions. Suppose the assumptions in Theorem \ref{th5.1} hold.  Let $\mathcal{S}_M$ be the set of all minimizers of $F$ in $W_M$ and  $(\r, {\bf v}, \Phi)(x, t)$ be an axi-symmetric weak solution of the Cauchy problem (\ref{1.1}), (\ref{initial}) with $\r(\cdot, t)\in L^1\cap L^{\g}$, $\r|{\bf v}|^2(\cdot, t)\in L^1$ and let $\nabla\Phi(\cdot, t)=-\nabla B\r(\cdot, t)\in L^2$.  If the total energy $E(t)$ is non-increasing with respect to $t$,
then for every $\epsilon>0$, there exists a number $\delta>0$ such that if
\begin{align} &\inf_{\tilde \r\in \mathcal{S}_M}\left[ d(\r_0, \tilde \r)+\f{1}{8\pi}||\nabla B\r_0-\nabla B\tilde \r||_2^2+ |d_1(\r_0, \tilde \r)|\right]\notag\\
&+\f{1}{2}\int \r_0(x)(|v^r_0|^2+|v^3_0|^2)(x)dx
<\delta,\end{align}
then for every $t>0$
\begin{align} &\inf_{\tilde \r\in \mathcal{S}_M}\left[d(\r(t), T^a\tilde \r)+\f{1}{8\pi}||\nabla B\r(t)-BT^a\tilde \r||_2^2+|d_1(\r(t), T^a\tilde \r)|\right]\notag\\
&+\f{1}{2}\int \r(x, t)(|v^r(x, t)|^2+|v^3(x, t)|^2)(x)dx
<\epsilon.
\end{align}
The proof of this follows exactly along the same line as that for Theorem \ref{th5.1}. \end{rem}

In order to prove Theorem \ref{th5.1}, we need several lemmas. First we have
\begin{lem}\label{lem5.2}
Suppose the angular momentum of the rotating star solutions satisfies  (\ref{L}), (\ref{L1}) and (2.13).  For any $\r(x)\in W_M$, if
\be\label{extra}
\lim_{r\to 0+}{L(m_\r(r)+m_{\tilde \r}(r))m_{\sigma}(r)}{r^{-2}}=0,
\ee
where $\sigma=\r-\tilde \r,$
then
\be\label{dd1} d_1(\r, \tilde \r)\ge 0,
\ee
where $d_1$ is defined by (\ref{d1}).
\end{lem}
\begin{proof} First, we introduce some notation. For an axi-symmetric  function $f(x)=f(r, z)$ ($r=\sqrt {x_1^2+x_2^2},\ z=x_3$ for $x=(x_1, x_2, x_3)$),  we let
\be \hat f(r)=2\pi r\int_{-\infty}^{+\infty} f(r, z)dz,\ee

\be\label{xd1} m_f(r)=\int_{\{x: \sqrt{x_1^2+x_2^2}\le r\}}f(x)dx=\int_0^r \hat f(s)ds,\ee
so that
\be\label{dx2}
m'_f(r)=\hat f(r).\ee
In order to show (\ref{dd1}), we let
\be \sigma(x)=(\r-\tilde \r)(x),\ee
and for $0\le \alpha\le 1$, we define
\begin{align}
Q(\alpha)&=\f{1}{2}\int\f{(\tilde \r+\alpha\sigma)(x) L(m_{\tilde \r+\alpha\sigma}(r(x)))-\tilde \r(x) L(m_{\tilde \r}(r(x)))}{r^2(x)}dx\notag\\
&-\alpha\int \int_{r(x)}^{\infty}s^{-3}L(m_{\tilde \r}(s))ds\sigma(x)dx.
\end{align}
 Then
\be\label{5.37} Q(0)=0,\ Q(1)=d_1(\r,\ \tilde \r). \ee
Since
\be m_{\tilde \r+\alpha\sigma}(r(x))=\int_0^{r(x)}2\pi s \int_{-\infty}^{+\infty} (\tilde \r+\alpha \sigma)(s, z)dzds,
\ee
we have
\be\f{d}{d\alpha} m_{\tilde \r+\alpha\sigma}(r(x))=\int_0^{r(x)}2\pi s \int_{-\infty}^{+\infty}\sigma(s, z)dzds= m_{\sigma}(r(x)).\ee
Therefore,
\begin{align}\label{dx1}
Q'(\alpha)&=\f{1}{2}\int\f{\sigma(x) L(m_{\tilde \r+\alpha\sigma}(r(x)))}{r^2(x)}dx\notag\\
&+\f{1}{2}\int\f{(\tilde \r+\alpha\sigma)(x) L'(m_{\tilde \r+\alpha\sigma}(r(x)))m_{\sigma}(r(x))}{r^2(x)}dx\notag\\
&-\int \int_{r(x)}^{\infty}s^{-3}L(m_{\tilde \r}(s))ds\sigma(x)dx,
\end{align}
and in view of (\ref{dx2}),
\vskip 0.2cm
\be\label{dx3} \f{d}{dr} L(m_{\tilde \r+\alpha\sigma}(r))=L'(m_{\tilde \r+\alpha\sigma}(r))(\ \hat{\tilde \r} +\alpha \hat \sigma)(r).\ee
\vskip 0.2cm
\noindent Therefore, by virtue of (\ref{dx3}) and (\ref{dx2}),   we obtain
\begin{align}\label{dx5'}
& \f{1}{2}\int\f{(\tilde \r+\alpha\sigma)(x) L'(m_{\tilde \r+\alpha\sigma}(r(x)))m_{\sigma}(r(x))}{r^2(x)}dx\notag\\
&=\f{1}{2}\int_0^{+\infty}(\ \hat{\tilde \r}+\alpha\hat\sigma)(r) L'(m_{\tilde \r+\alpha\sigma}(r))m_{\sigma}(r)r^{-2}dr\notag\\
&=\f{1}{2}\int_0^{+\infty}\f{d}{dr}[L(m_{\tilde \r+\alpha\sigma}(r))]m_{\sigma}(r)r^{-2}dr.\end{align}
\vskip 0.2cm
\noindent For $0\le \alpha\le 1$, since (cf. (2.13))  $L'(m)\ge 0$, we have
\be \label{good1}
L(m_{\tilde \r+\alpha\sigma}(r))\le L(m_{\tilde \r+\rho}(r)).\ee
This, together with (\ref{extra}), implies
\be\label{good2} \lim_{r\to 0+}L(m_{\tilde \r+\alpha\sigma}(r))m_{\sigma}(r)r^{-2}=0.\ee
Moreover, since $m_{\sigma}(+\infty)=\int\sigma(x)dx=\int (\r-\tilde \r)(x)=0$ and $$\lim_{r\to \infty}L(m_{\tilde \r+\alpha\sigma}(r)=L(M),$$
we have \be\label{good3} \lim_{r\to \infty}L(m_{\tilde \r+\alpha\sigma}(r))m_{\sigma}(r)r^{-2}=0.\ee
\vskip 0.2cm
\noindent It follows from (\ref{dx5'}), (\ref{good2}), (\ref{good3}) and integration by parts that
\begin{align}\label{dx5}
& \f{1}{2}\int\f{(\tilde \r+\alpha\sigma)(x) L'(m_{\tilde \r+\alpha\sigma}(r(x)))m_{\sigma}(r(x))}{r^2(x)}dx\notag\\
&=-\f{1}{2}\int_0^{+\infty}\hat \sigma (r)L(m_{\tilde \r+\alpha\sigma}(r))m_{\sigma}(r)r^{-2}dr\notag\\
&+\int_0^{+\infty}L(m_{\tilde \r+\alpha\sigma}(r))m_{\sigma}(r)r^{-3}dr.
\end{align}

Since
\be \int_0^{+\infty}\hat \sigma (r)L(m_{\tilde \r+\alpha\sigma}(r))m_{\sigma}(r)r^{-2}dr=\int\f{\sigma(x) L(m_{\tilde \r+\alpha\sigma}(r(x)))}{r^2(x)}dx,\ee
and \be \int \int_{r(x)}^{\infty}s^{-3}L(m_{\tilde \r}(s))ds\sigma(x)dx=\int_0^{+\infty}\hat \sigma(r)\int_{r}^{\infty}s^{-3}L(m_{\tilde \r}(s))dsdr,\ee

 (\ref{dx1}) and (\ref{dx5}) imply
\begin{align}\label{dx6}
Q'(\alpha)&=\int_0^{+\infty}L(m_{\tilde \r+\alpha\sigma}(r))m_{\sigma}(r)r^{-3}dr\notag\\
&-\int_0^{+\infty}\hat \sigma(r)\int_{r}^{\infty}s^{-3}L(m_{\tilde \r}(s))dsdr.\end{align}
Using (\ref{xd1}), we have $m_\sigma(r)=\int_0^r\hat \sigma(s)ds$, so substituting this into the first term  in (\ref{dx6})
and interchanging the order of integration gives
 \begin{align}\label{dx7}&\int_0^{+\infty}L(m_{\tilde \r+\alpha\sigma}(r))m_{\sigma}(r)r^{-3}dr\notag\\
 &=\int_0^{+\infty}\int_0^r r^{-3}L(m_{\tilde \r+\alpha\sigma}(r))\hat \sigma(s)dsdr\notag\\
 &=\int_0^{+\infty}\hat \sigma(s)\int_s^{+\infty}r^{-3}L(m_{\tilde \r+\alpha\sigma}(r))drds\notag\\
 &=\int_0^{+\infty}\hat \sigma(r)\int_r^{+\infty}s^{-3}L(m_{\tilde \r+\alpha\sigma}(s))dsdr.\end{align}
 Hence (\ref{dx6}) and (\ref{dx7}) yield
 \be\label{dx8}
Q'(\alpha)=\int_0^{+\infty}\hat \sigma(r)\int_{r}^{\infty}s^{-3}(L(m_{\tilde \r+\alpha\sigma}(s))-L(m_{\tilde \r}(s)))dsdr,\ee
and therefore
\be\label{dx9} Q(0)=Q'(0)=0.\ee
Differentiating (\ref{dx8}) again, we obtain
\be\label{dx91}
\f{d^2Q(\alpha)}{d\alpha^2}=\alpha\int_0^{+\infty}\hat \sigma(r)\int_{r}^{\infty}s^{-3}L'(m_{\tilde \r+\alpha\sigma}(s))m_\sigma(s)dsdr,\ee
and interchanging the order of integration gives
\be\label{dx10}
\f{d^2Q(\alpha)}{d\alpha^2}=\alpha\int_0^{+\infty}s^{-3}\int_0^s\hat \sigma(r)dr L'(m_{\tilde \r+\alpha\sigma}(s))m_\sigma(s)ds.\ee
Noting  that $\int_0^s\hat \sigma(r)dr=m_\sigma(s)$, we obtain
\be\label{dx101}
\f{d^2Q(\alpha)}{d\alpha^2}=\alpha\int_0^{+\infty}s^{-3}L'(m_{\tilde \r+\alpha\sigma}(s))(m_\sigma(s))^2ds.\ee
Therefore, if $L'(m)\ge 0$ for $0\le m\le M$, then
\be\label{dx11}\f{d^2Q(\alpha)}{d\alpha^2}\ge 0,\  for \ 0\le\alpha\le 1.\ee
This, together with (\ref{dx9})and (\ref{5.37}),  yields $d_1(\r, \tilde \r)=Q(1)\ge 0.$
 \end{proof}

\begin{lem}\label{lem5.3} Let $(\r, {\bf v})$ be a solution of the Cauchy problem (\ref{1.1}), (\ref{initial}) as stated in Theorem 3.1, then
\begin{align}\label{ed}
&E(\r, {\bf v})(t)-F(\tilde \r)\notag\\
&=d(\r(t),\tilde \r)+d_1(\r(t), \tilde \r)
-\f{1}{8\pi}||\nabla (B\r(\cdot, t)-B\tilde \r)||_2^2\notag\\
&+\f{1}{2}\int\r (|v^r|^2+|v^3|^2)(x, t)dx.\end{align}\end{lem}
\begin{proof}
From A1)-A3), for any $x\in G_t$ we have
\be\label{jj1}j^2(x, t)=j_0^2(\xi_{-t}(x)),\ee
(see (3.26)).  In view of  (3.22) and (3.27),
\be\label{jj2}
j^2(x, t)=j_0^2(\xi_{-t}(x))=L(m_{\r_0}(r(\xi_{-t}(x))),
\ee
for $x\in G_t$.
This, together with (3.21), yields
\be\label{jj3}
j^2(x, t)=L(m_{\r(t)}(r(x))),\qquad x\in G_t.\ee
Therefore, by (\ref{en}), we have
\begin{align}\label{jj4}
E(\r(t), {\bf v}(t))
&=\int A(\r)(x, t)dx+\frac{1}{2}\int \f{\r(x, t) L(m_{\r(t)}(r(x))}{r^2(x)}dx\notag\\
&-\frac{1}{8\pi}\int|\nabla B\r|^2(x, t)dx+\frac{1}{2}\int \r(|v^r|^2+|v^3|^2)(x, t)dx.\end{align}
Here we have used the fact that
$$\int \f{\r(x, t) L(m_{\r(t)}(r(x))}{r^2(x)}dx=\int_{G_t} \f{\r(x, t) L(m_{\r(t)}(r(x))}{r^2(x)}dx,$$
which holds because $D_t=\{x\in R^3: \r(x, t)>0\}$, $G_t\subset D_T$  and $meas(D_t-G_t)=0.$
It follows from (2.5) and (\ref{jj4}) that
\begin{align}\label{jj5}
&E(\r, {\bf v})(t)-F(\tilde \r)\notag\\
&=\int
(A(\rho)(x,t)-A(\tilde \r)(x))dx\notag\\
&+\f{1}{2}\int\f{\r(x, t) L(m_{\r(t)}(r(x))-\tilde \r(x) L(m_{\tilde \r}(r(x))}{r^2(x)}dx\notag\\
&-\f{1}{8\pi}(||\nabla B\r(x, t)||^2-||\nabla B\tilde \r||_2^2)\notag\\
&+\f{1}{2}\int\r (|v^r|^2+|v^3|^2)(x, t)dx.\end{align}
On the other hand,
\begin{align}\la{jj6}
&-\f{1}{8\pi}(||\nabla B\r(\cdot, t)||_2^2-||\nabla B\tilde \r||_2^2)\notag\\
&=-\f{1}{8\pi}||\nabla (B\r(\cdot, t)-\nabla B\tilde \r)||_2^2-\f{1}{4\pi}\int \nabla B\tilde \r(x)\cdot (\nabla B\r(x, t)-\nabla B\tilde \r(x))dx.\end{align}
Noting that $\Delta (B\r-B\tilde \r)=-4\pi (\r-\tilde \r)$, and integrating by parts (this is legitimate, cf. \cite{rein1}) gives,
\begin{align}\label{jj8}
&-\f{1}{4\pi}\int \nabla B\tilde \r(x)\cdot (\nabla B\r(x, t)-\nabla B\tilde \r(x))dx\notag\\
&=\f{1}{4\pi}\int  B\tilde \r(x)(\Delta B\r(x, t)-\Delta B\tilde \r(x))dx\notag\\
&=\int  B\tilde \r(x) (\r(x, t)- \tilde \r(x))dx.\end{align}
By (\ref{jj5})-(\ref{jj8}), and noting (\ref{d1}), we have
\begin{align}\label{jj9}
& E(\r, {\bf v})(t)-F(\tilde \r)\notag\\
&=\int \left(A(\rho)-A(\tilde \r)+(\r-\tilde \r)\{\int_{r(x)}^{\infty}\f{L(m_{\tilde \r}(s))}{s^3}ds-B\tilde \r\}\right)dx\notag\\
&+d_1(\r(t), \tilde \r)-\f{1}{8\pi}(||\nabla (B\r(x, t)- B\tilde \r)||_2^2\notag\\
&+\f{1}{2}\int\r (|v^r|^2+|v^3|^2)(x, t)dx.\end{align}
Since $\r(\cdot, t)\in W_M$,  $\int \r(x, t)dx=\int\tilde \r(x)dx=M.$ Thus $\int \lambda(\r(x, t)-\tilde \r(x))dx=0.$
Therefore, the first term in (\ref{jj9}) is the same as $d(\r(t), \tilde \r)$ defined by (3.29). This completes the proof of the lemma.
\end{proof}

\noindent We are now in a position to prove Theorem 3.1.

\vskip 0.3cm

\noindent {\it Proof of Theorem 3.1.} Assume  the theorem is false. Then there exist $\epsilon_0>0$, $t_n>0$ and initial data
$\r_n(x,0)\in W_M$ and ${\bf v}_n(x, 0)$ such that for all $n\in \mathbb{N}$,
\begin{align}\label{bn1} &d(\r_n(0), \tilde \r)+d_1(\r_0, \tilde \r)+\f{1}{8\pi}||\nabla B\r_n(0)-\nabla B\tilde \r||_2^2\notag\\
&+\f{1}{2}\int \r_n(x, 0)(|v_n^r(x, 0)|^2+|v_n^3(x, 0)|^2)(x)dx<\f{1}{n},
\end{align}
but for any $a\in \RR$,
\begin{align}\label{bn2} &d(\r_n(t_n), T^a\tilde \r)+ d_1(\r(t), T^a\tilde \r)+\f{1}{8\pi}||\nabla B\r_n(t_n)-\nabla BT^a\tilde \r||_2^2\notag\\
&+\f{1}{2}\int \r_n(x, t_n)(|v_n^r(x, t_n)|^2+|v_n^3(x, t_n)|^2)(x)dx\ge \epsilon_0.
\end{align}
By (\ref{ed}) and (\ref{bn1}), we have
\be\label{bn3}
\lim_{n\to \infty}E(\r_n(0), {\bf v}_n(0))=F(\tilde \r).
\ee
Since $E(\r_n(t), {\bf v}_n(t))$ is non-increasing in time,
\be
 \lim_{n\to \infty}\sup F(\r_n(t_n))\le \lim_{n\to \infty} E(\r_n(t_n), {\bf v}_n(t_n))
 \le \lim_{n\to \infty} E(\r_n(0), {\bf v}_n(0))=F(\tilde \r).\ee
 (The first inequality holds because we have,  similar to (3.71),
 $$E(\r, {\bf v})(t)-F(\r(t))=\f{1}{2}\int \r(|v^r|^2+|v^3|^2)(x, t)dx\ge 0,\ t\ge 0.)$$
 Therefore $\{\r_n(\cdot, t_n)\}\subset W_M$ is a minimizing sequence for the functional $F$.
 We  apply
 Theorem \ref{aa} to conclude that there exists a sequence $\{a_n\}\subset \RR$ such that up to a subsequence,
 \be ||\nabla(B\r_n(t_n)-BT^{a_n}\tilde \r)||_2\to 0, \ee
 as $n\to \infty$; this is where we use the assumption that the minimizer is unique up to a vertical shift. Note also that for any $\r\in W_M$ and
 $a\in \RR$,
 \be\
 \begin{cases}& ||\nabla B(T^a\r)-\nabla B\tilde \r||_2= ||\nabla B(\r)-\nabla BT^{-a}\tilde \r||_2,\\
   &d(T^a \r, \tilde \r)=d(\r, T^{-a}\tilde \r),\  and\  d_1(T^a \r, \tilde \r)=d_1(\r, T^{-a}\tilde \r).\end{cases}\ee
 Thus, by (\ref{ed}), the fact that the energy is non-increasing in time,  and $F(T^a\r)=F(\r)$,  we have for any $\r\in W_M$ and $a\in \RR$,
 \begin{align}
&E(\r_n(t_n), {\bf v}_n(t_n))-F(T^{a_n}\tilde \r)\notag\\
&=d(\r_n(t_n),T^{a_n}\tilde \r)+d_1(\r(t_n), T^{a_n}\tilde \r)\notag\\
&-\f{1}{8\pi}||\nabla(B\r_n(t_n)-BT^{a_n}\tilde \r)||_2^2\notag\\
&+\f{1}{2}\int\r_n (|v_n^r|^2+|v_n^3|^2)(x, t_n)dx\notag\\
&\le E(\r_n(0), {\bf v}_n(0))-F(T^{a_n}\tilde \r)\notag\\
&=E(\r_n(0), {\bf v}_n(0))-F(\tilde \r)\to 0,\end{align}
as $n\to\infty$.
Since $$||\nabla B\r_n(t_n)-\nabla BT^{a_n}\tilde \r||_2\to 0,$$ as $n\to \infty$,  $d(\r_n(t_n),\tilde \r)\ge 0$ (cf. (3.31)) and
$d_1(\r(t_n), \tilde \r)\ge 0$ (cf. A4) and (3.37)), we have
\begin{align}&d(\r_n(t_n),T^{a_n}\tilde \r)+d_1(\r(t_n), T^{a_n}\tilde \r)\notag\\
&+\f{1}{8\pi}||\nabla(B\r_n(t_n)-T^{a_n}B\tilde \r)||_2^2\notag\\
&+\f{1}{2}\int\r_n (|v_n^r|^2+|v_n^3|^2)(x, t_n)dx\to 0,
\end{align}
as $n\to \infty$.
This contradicts (\ref{bn2}), and completes the proof.  \\
\section{Stability of General Entropy Solutions }
In this section, we shall obtain  a stability theorem for general
entropy weak solutions.   We
begin with the definition of entropy weak solution. \vskip
0.2cm \noindent  {\bf Definition 4.1}. A weak solution (defined in
Section 3) on $\RR^3\times [0, T]$  is called an {\it entropy weak
solution} of (\ref{1.1}) if it satisfies the following ``entropy
inequality'':
\begin{equation}\label{entropy1}
\partial_t \eta +\sum_{j=1}^{3}\partial_{x_j}q_j+\rho \sum_{j=1}^{3}\eta_{m_j}\Phi_{x_j}\le 0,\end{equation} in the sense of distributions; i.e.,
\be\label{entro} \int_0^T\int_{\RR^3}\left(\eta \beta_t+{\bf q}\cdot
\nabla \beta-\rho
\sum_{j=1}^{3}\eta_{m_j}\Phi_{x_j}\beta\right)dxdt+\int_{\RR^3}\beta(x,
0)\eta(x, 0)dx\ge 0,\ee for any nonnegative Lipschitz continuous  test function $\beta$ with compact support in $[0, T)\times \RR^3$. Here the ``entropy'' function $\eta$ and ``entropy flux" functions $q_j$ and
${\bf q}$,  are defined by
\begin{equation}
\begin{cases}
&\eta=\frac{|{\bf m}|^2}{2\rho}+\r\int_0^\r\f{p'(s)}{s^2}ds=\frac{|{\bf m}|^2}{2\rho}+\f{\r^\g}{\g-1},\\
&q_j=\frac{|{\bf
m}|^2m_j}{2\rho^2}+m_j\int_0^{\rho}\frac{p'(s)}{s}ds=\frac{|{\bf m}|^2}{2\rho}+\f{\g\r^\g}{\g-1}\quad  (j=1, 2, 3),\\
&{\bf q}=(q_1,\ q_2, \ q_3). \end{cases}
\end{equation}

\begin{rem} The inequality (\ref{entropy1}) is motivated by the second law of thermodynamics (\cite{lax}), and plays an important role in shock wave theory (\cite{smoller}). For smooth solutions, the inequality in (\ref{entropy1}) can be replaced by equality. \end{rem}

For a general entropy weak solution, our stability result is given by the following theorem:

\begin{thm}\label{thm5.1} Suppose $1<\gamma\le 2$. Let $(\rho, {\bf m}, \Phi)(x, t)$  ($t\in [0, T]$, $x\in\RR^3$) with $(\r, {\bf m})\in L^{\infty}( \RR^3\times [0, T])$,  be a weak solution of (\ref{1.1}) satisfying the entropy condition (\ref{entropy1}) and let $(\bar\rho, \bar{ \bf m}, \bar \Phi)(x, t)$, $t\in [0, T],\ x\in \RR^3$ be  any  solution of (\ref{1.1})
satisfying $(\bar\rho, \bar {\bf m})\in W^{1, \infty}_{loc}(\RR^3\times [0, T]).$
 Assume
\be Z(T)=: \sup_{0\le t \le T}(||\r(\cdot, t)||_{\infty}(||\r(\cdot, t)||_{\infty}+||\bar \r(\cdot, t)||_{\infty})^{2-\gamma}(Vol  S(t))^{2/3}+||\nabla_x\bar{\bf v}(\cdot, t)||_{\infty})<+\infty,\ee
and  \be\label{v13}\f{{\bf m}}{ \r}, \f{\bar {\bf m}}{\bar \r}\in L^{\infty}(\RR^3\times [0, T]).\ee
 where $S(t)=Supp |\rho-\bar \r|(\cdot, t)$.  Then there is a constant $C(T)$ depending on $T$ and $Z(T)$ such that
 \begin{equation} Y(t)\le C(T) Y(0),\qquad 0\le t\le T,\end{equation}
 where
 $$Y(t)= D(\rho, \bar \rho)(t)+||\sqrt{\r}(\nabla \Phi-\nabla \bar\Phi)||_2^2(t)+\int \rho(x, t)|{\bf v}-\bar {\bf v}|^2(x, t) dx, $$
 $$\Phi=-B\r, \bar \Phi=-B \bar\r,$$
 and
 $$D(\rho,\ \bar \rho)= \int\f{p(\rho)-p(\bar\rho)-p'(\bar\rho)(\r-\bar\r)}{\g-1}dx.$$
\end{thm}
\begin{rem}  The function $(\bar \r, \bar{\bf m},  \bar \Phi)$ in the theorem could be, but is not necessarily,  a rotating star solution.
\end{rem}
\begin{rem}
For $1<\g\le 2$, it is easy to see
$$D(\rho,\ \bar \rho)\ge C||\r-\bar \r||_2^2, $$
for some constant $C>0$ if $\rho, \bar\rho \in L^{\infty}(\RR^3\times [0, T])$.
\end{rem}

\noindent {\it Proof of Theorem \ref{thm5.1}}\\
\noindent Letting $U=(\rho, {\bf m})^{\mathrm T}$ with ${\bf m}=(m_1, m_2, m_3)=\rho{\bf v}$ and $\bar U=(\bar \rho, \bar {\bf m})^{\mathrm T}$,
we can write  system (\ref{1.1}) as
\begin{equation}\label{1.2}
\begin{cases}
&U_t+\sum_{j=1}^{3}F_j(U)_{x_j}=-\rho \nabla \Phi,\\
 &\Delta \Phi=4\pi \rho.
\end{cases}
\end{equation}
Here the flux functions $F_j(U)$  are given by
\begin{equation}
\begin{cases}
& F_1(U)=\left(m_1, p(\rho)+\frac{m_1^2}{\rho},\
\frac{m_1m_2}{\rho},\ \frac{m_1m_3}{\rho}\right)^{\mathrm T},\\ &
F_2(U)=\left(m_2, \frac{m_1m_2}{\rho}, p(\rho)+\frac{m_2^2}{\rho},\
\frac{m_2m_3}{\rho}\right)^{\mathrm T},\\ &F_3(U)=\left(m_3,
\frac{m_1m_3}{\rho}, \frac{m_2m_3}{\rho},\
p(\rho)+\frac{m_3^2}{\rho}\right)^{\mathrm T}.\end{cases}\end{equation}
 The entropy and entropy fluxes $\eta$ and ${\bf q}$ are as in (4.3) and satisfy
\be\label{entropy2}
\nabla q_j(U)=\nabla \eta (U)\nabla F_j(U), \qquad j=1,\ 2, 3,\ee as is easily verifiable.
Since $U$ is an entropy weak solution
\begin{equation}\label{entrox1}
\partial_t \eta(U) +\sum_{j=1}^{3}\partial_{x_j}q_j(U)+\rho \sum_{j=1}^{3}\eta_{m_j}(U)\Phi_{x_j}\le 0,\end{equation} in the sense of distributions.
Because $\bar U\in W_{loc}^{1,\ \infty}$ is a weak solution of (\ref{1.1}), we have
\begin{equation}\label{entrox2}
\partial_t \eta(\bar U) +\sum_{j=1}^{3}\partial_{x_j}q_j(\bar U)+\rho \sum_{j=1}^{3}\eta_{m_j}(\bar U)\bar \Phi_{x_j}= 0.\end{equation}
We define the relative entropy-entropy flux pairs by
\begin{equation}\label{re}
\begin{cases}&\eta^*(U, \bar U)=\eta(U)-\eta(\bar U)-\nabla \eta(\bar U)(U-\bar
U),\\
&q_j^*(U, \bar U)=q_j(U)-q_j(\bar U)-\nabla \eta(\bar
U)(F_j(U)-F_j(\bar U))\ (j=1, 2, 3).\end{cases}\end{equation}

 Using (\ref{entrox1}) and (\ref{entrox2}) gives
\begin{align}\label{entropy}
&\partial_t \eta^*+\sum_{j=1}^3 \partial_{x_j}q^*_j\notag\\
&=(\partial_t \eta(U)+\sum_{j=1}^3 \partial_{x_j}q_j(U))-(\partial_t
\eta(\bar U)+\sum_{j=1}^3 \partial_{x_j}q_j(\bar U))\notag\\
&-\nabla^2\eta(\bar U)\{(\bar U_t, U-\bar
U)+(\sum_{j=1}^3\partial_{x_j}\bar U, F_j(U)-F_j(\bar U))\}\notag\\
&-\nabla \eta(\bar U)\{(U-\bar
U)_t+\sum_{j=1}^3\partial_{x_j}(F_j(U)-F_j(\bar U))\}\notag\\
 &\le
(\nabla \eta(U)-\nabla\eta(\bar U))R-\nabla^2 \eta (\bar U)(\bar R,
U-\bar U)\notag\\
&-\nabla^2 \eta (\bar U)\sum_{j=1}^3\left(\partial_{x_j}\bar U,\
F_j(U)-F_j(\bar U)-F_j'(\bar U)(U-\bar U)\right),\end{align}
in the sense of distributions,  where
\be R=(0, -\rho \nabla \Phi)^{\mathrm T},{~\rm and~} \bar R=(0, -\bar \rho  \nabla
\bar\Phi)^{\mathrm T}.\ee
It is easy to check that \begin{align} &(\nabla \eta(U)-\nabla\eta(\bar
U))R-\nabla^2 \eta (\bar U)(\bar R, U-\bar U)\notag\\
&=-\rho({\bf v}-\bar {\bf v})\cdot(\nabla \Phi-\nabla \bar
\Phi),\end{align}
so that
\begin{align}
&\partial_t \eta^*+\sum_{j=1}^3 \partial_{x_j}q^*_j\notag\\
&\le -\rho({\bf v}-\bar {\bf v})\cdot(\nabla \Phi-\nabla \bar \Phi)\notag\\
&-\nabla^2 \eta (\bar U)\sum_{j=1}^3\left(\partial_{x_j}\bar U,\
F_j(U)-F_j(\bar U)-F_j'(\bar U)(U-\bar U)\right),\end{align}
in the sense of distributions. That is, for any nonnegative, Lipschitz continuous test function
$\psi$ on $\RR^3\times [0, T)$, with compact support, we have
\begin{align}\label{121}
&\int_0^T\int_{\RR^3}\left(\partial_t\psi \eta^*+\sum_{j=1}^3 \partial_{j}\psi q^*_j\right)dxdt+\int_{\RR^3}\psi(x, 0)\eta^*(x, 0)dx\notag\\
&\ge \int_0^T\int_{\RR^3}\psi \rho({\bf v}-\bar {\bf v})\cdot(\nabla \Phi-\nabla \bar \Phi)dxdt\notag\\
&+\int_0^T\int_{\RR^3}\psi\nabla^2 \eta (\bar U)\sum_{j=1}^3\left(\partial_{x_j}\bar U,\
F_j(U)-F_j(\bar U)-F_j'(\bar U)(U-\bar U)\right)dxdt.\end{align}
A calculation gives
\begin{equation}\la{qua}
\nabla^2\eta(U)=\begin{pmatrix} &\frac{m^2}{\rho^3}+\frac{p''(\rho)}{\gamma-1} &  -\frac{m_1}{\rho^2} &   -\frac{m_2}{\rho^2} &   -\frac{m_3}{\rho^2}\\
&-\frac{m_1}{\rho^2} & \frac{1}{\rho} & 0 & 0\\
&-\frac{m_2}{\rho^2} & 0 & \frac{1}{\rho} &
0\\
&-\frac{m_3}{\rho^2} & 0 & 0 & \frac{1}{\rho}
\end{pmatrix},\end{equation}
and also  \be\label{eta}
\eta^*=\f{p(\r)-p(\b\r)-p'(\b\r)(\r-\b\r)}{\g-1}+\f{1}{2}\r|\v
v-\bar{\v{v}}|^2. \ee
So,  for $1<\gamma\le 2$,
 \be\label{eta2}
 \eta^*\ge c_1 (||\r(\cdot, t)||_{\infty}+||\bar \r(\cdot, t)||_{\infty})^{\gamma-2}(\rho-\bar \rho)^2 +\f{1}{2}\rho|{\bf v}-\bar {\bf  v}|^2\ge 0, \ee
 for some positive constant $c_1$.

A further calculation yields, using (\ref{qua}),
\begin{align}\label{122} & \nabla^2 \eta (\bar
U)\sum_{j=1}^3\left(\partial_{x_j}\bar U,\ F_j(U)-F_j(\bar
U)-F_j'(\bar U)(U-\bar U)\right)\notag\\
&=\{p(\rho)-p(\bar\rho)-p'(\bar\rho)(\rho-\bar\rho)\}\sum_{j=1}^3\partial_j\bar
{v}_j\notag\\
&+\f{1}{2}\sum_{i, j=1}^3\rho(v_i-\bar v_i)(v_j-\bar
v_j)(\partial_j\bar v_i+\partial_i\bar v_j).
\end{align}
Here and in the following, we use the notation:
$$\partial_j=\f{\partial}{\partial{x_j}},\qquad j=1,\ 2,\ 3.$$
Therefore, by (\ref{eta}) and (\ref{122}), we have
\begin{align}\label{123} & |\nabla^2 \eta (\bar
U)\sum_{j=1}^3\left(\partial_{x_j}\bar U,\ F_j(U)-F_j(\bar
U)-F_j'(\bar U)(U-\bar U)\right)|(x, t)\notag\\
& \le C ||\nabla_x \bar {\bf v}(\cdot, t)||_{\infty}\eta^*(x, t),
\end{align}
for $x\in\RR^3$,  $t\in [0, T)$ and some constant $C>0$.
Thus, (\ref{121})-(\ref{123}) yield
\begin{align}\label{124}
&\int_0^T\int_{\RR^3}\left(\partial_t\psi \eta^*+\sum_{j=1}^3 \partial_{j}\psi q^*_j\right)dxdt+\int_{\RR^3}\psi(x, 0)\eta^*(x, 0)dx\notag\\
&\ge \int_0^T\int_{\RR^3}\psi \rho({\bf v}-\bar {\bf v})\cdot(\nabla \Phi-\nabla \bar \Phi)dxdt\notag\\
&- C \sup_{0\le t\le T}||\nabla_x \bar {\bf v}(\cdot, t)||_{\infty}\int_0^T\int_{\RR^3}\psi\eta^*(x, t)dxdt.
\end{align}
Using (\ref{v13}),  it is easy to see that there exists a positive constant $\Lambda$, which may depend on $T$,   such that
\be\label{125}
(\sum_{j=1}^3|q^*_j|^2)^{1/2}(x, t)\le \Lambda \eta^*(x, t), \qquad (x, t)\in \RR^3\times [0, T].
\ee
For  fixed $L>0$, $t\in (0, T)$ and small  $\epsilon>0$,  we consider the test function $\psi (x, \tau)=\varsigma(x, \tau)\vartheta(\tau)$ defined by
\be\label{test1}
\vartheta (\tau)=\begin{cases} &1, \qquad\qquad\qquad 0\le \tau<t\\
&\f{1}{\epsilon}(t-\tau)+1, \quad t\le \tau<t+\epsilon\\
& 0, \qquad\qquad\qquad t+\epsilon\le \tau<T,
\end{cases}
\ee
\be\label{test2}
\varsigma(x, \tau)=\begin{cases} &1, \qquad\qquad\qquad\qquad\qquad\qquad  (x, \tau)\in R_1\\
&\f{1}{\epsilon}[L+\Lambda(t-\tau)-|x|]+1,\qquad  (x, \tau)\in R_2\\
&0, \qquad\qquad\qquad\qquad\qquad\qquad(x, \tau)\in R_3,
\end{cases}
\ee
where $$R_1=\{(x, \tau): 0\le \tau<T,\  0\le |x|<L+\Lambda(t-\tau)\},$$
$$R_2=\{(x, \tau): 0\le \tau<T, L+\Lambda(t-\tau)\le |x|<L+\Lambda(t-\tau)+\epsilon\},$$
$$R_3=\{(x, \tau): 0\le \tau<T, \ |x|>L+\Lambda(t-\tau)+\epsilon\},$$
and $\Lambda$ is the constant given in (\ref{125}). Substituting this in (\ref{124}),  a straightforward calculation yields,
\begin{align}\label{test3}
&\f{1}{\epsilon}\int_{t}^{t+\epsilon}\int_{|x|<L}\eta^*(x, \tau)dxd\tau\notag\\
&\le \int_{|x|<L+\Lambda t}\eta^*(x, 0)dx\notag\\
&-\frac{1}{\epsilon}\int_0^t\int_{L+\Lambda(t-\tau)\le |x|<L+\Lambda(t-\tau)+\epsilon}\left\{\Lambda\eta^*+\sum_{j=1}^3\f{x_j}{|x|}q^*_j\right\}dxd\tau\notag\\
&-\int_0^t\int_{|x|<L+\Lambda(t-\tau)}\rho({\bf v}-\bar {\bf v})\cdot(\nabla \Phi-\nabla \bar \Phi)dxd\tau\notag\\
&+C\sup_{0\le \tau \le T}||\nabla_x \bar {\bf v}(\cdot, \tau)||_{\infty}\int_0^t\int_{|x|<L+\Lambda(t-\tau)}\eta^*(x, \tau)dxd\tau+O(\epsilon).
\end{align}
\vskip 0.2cm
\noindent The second term on the right-had side of (\ref{test3}) is negative in view of (\ref{125}), together with Cauchy-Schwartz inequality. Letting $\epsilon\to 0^+$ in (\ref{test3}) gives
 \begin{align}\label{test4}
&\int_{|x|<L}\eta^*(x, t)dx\notag\\
&\le \int_{|x|<L+\Lambda t}\eta^*(x, 0)dx\notag\\
&-\int_0^t\int_{|x|<L+\Lambda(t-\tau)}\rho({\bf v}-\bar {\bf v})\cdot(\nabla \Phi-\nabla \bar \Phi)dxd\tau\notag\\
&+C\sup_{0\le \tau \le T}||\nabla_x \bar {\bf v}(\cdot, \tau)||_{\infty}\int_0^t\int_{|x|<L+\Lambda(t-\tau)}\eta^*(x, \tau)dxd\tau.
\end{align}
We now let $L\to +\infty$ in (\ref{test4}) to get
\begin{align}\label{test5}
&\int \eta^*(x, t)dx\notag\\
&\le \int \eta^*(x, 0)dx\notag\\
&-\int_0^t\int\rho({\bf v}-\bar {\bf v})\cdot(\nabla \Phi-\nabla \bar \Phi)dxd\tau\notag\\
&+C\sup_{0\le \tau \le T}||\nabla_x \bar {\bf v}(\cdot, \tau)||_{\infty}\int_0^t\int\eta^*(x, \tau)dxd\tau.
\end{align}
The second term on the right hand side can be estimated as follows. By Cauchy-Schwartz inequality, we have
 \begin{align}&\label{x9}|\int\rho({\bf v}-\bar {\bf v})\cdot (\nabla \Phi-\bar \nabla \Phi)(x, \tau)dx \notag\\
 &\le \f{1}{2}\int\rho|{\bf v}-\bar {\bf v}|^2(x, \t)dx +\f{1}{2}\int\rho|\nabla \Phi-\bar \nabla \Phi|^2(x, \t)dx.\end{align}
 Applying Lemma \ref{lem2.2}, we obtain
 \begin{align}\label{x10}&\int\rho|\nabla \Phi-\bar \nabla \Phi|^2(x, t)dx\notag\\
 &\le ||\rho(\cdot,\t)||_{\infty}||\nabla (\Phi-\bar \Phi)(\cdot, \t||_2^2\notag\\
&\le C ||\rho(\cdot,\t)||_{\infty}\left(\int|\rho-\bar \rho|^{4/3}(x, t)dx\right)\left(\int|\rho-\bar \rho|^{4/3}(x, \t)dx\right)^{2/3}\notag\\
&=C ||\rho(\cdot,t)||_{\infty}\left(\int_{S(\t)}|\rho-\bar \rho|^{4/3}(x, t)dx\right)\left(\int_{S(\t)}|\rho-\bar \rho|^{4/3}(x, \t)dx\right)^{2/3},
\end{align}
where $$S(\t)=supp |\rho-\bar \rho|(\cdot, \t).$$
It follows from H${\rm \ddot{o}}$lder's inequality that
\be\label{x11}\int_{S(t)}|\rho-\bar \rho|^{4/3}(x, t)dx\le \left(\int_{S(\t)}|\rho-\bar \rho|^2(x, \t)dx\right)^{2/3}(vol S(\t))^{1/3},\ee
and
\be\label{x12}\left(\int_{S(t)}|\rho-\bar \rho|(x, \t)dx\right)^{2/3}\le \left(\int_{S(t)}|\rho-\bar \rho|^2(x, \t)dx\right)^{1/3}(vol S(\t))^{1/3}.\ee
Then using (\ref{x10})-(\ref{x12}) we obtain
 \be \label{x13}\int\rho|\nabla \Phi-\bar \nabla \Phi|^2(x, \t)dx\le C ||\rho(\cdot, \t)||_{\infty}(||\r(\cdot, \t)||_{\infty}+||\bar \r(\cdot, \t)||_{\infty})^{2-\g}||(\r-\bar \r)(\cdot, \t)||_2^2(Vol S(t)))^{2/3}.\ee
  \ In view of (\ref{eta2}), (4.29),  (4.30) and (4.34), we have
  \be
 \int \eta^*(x, t)dx\le \int \eta^*(x, 0)dx + C Z(T)\int_0^t \int\eta^*(x, \t)dxd\t,\ee
 for $0\le t\le T$, where
$$Z(T)=: sup_{0\le t \le T}(||\r(\cdot, t)||_{\infty}(||\r(\cdot, t)||_{\infty}+||\bar \r(\cdot, t)||_{\infty})^{2-\gamma}
 (Vol S(t))^{2/3}+||\nabla_x\bar{\bf v}(\cdot, t)||_{\infty}).$$
 Then (4.6) follows from Gronwall's inequality applied to (4.35) and using (4.19) and (4.20). This completes the proof of Theorem 4.1.

\section{ Uniform A Priori Estimates}

The theorem proved in this section gives  a uniform a priori estimate for the entropy weak solution defined in (4.2) of the Cauchy problem (\ref{1.1}) and (\ref{initial}).
As we shall see, this estimate justifies some assumptions made in Section 3 and should be useful for obtaining the existence of global weak solutions for the Cauchy problem.

\begin{thm}\label{thm4.2}
  If $(\r, {\bf m})\in L^{\infty}([0, T];  L^1(\RR^3))$ satisfies the first equation in (1.1) in the sense  of distributions, then \be\label{6.1}\int_{\RR^3} \r(x, t)dx=\int_{\RR^3}\r(x, 0)dx=:M, \qquad 0<t<T.\ee
 Let $(\rho, {\bf m}, \Phi)$ be a weak solution defined in Definition 3.1.  Suppose $(\rho, {\bf m}, \Phi)$ satisfies the entropy condition (4.2), $\rho\in  L^{\infty}([0, T];  L^1(\RR^3))\cap L^{\infty}([0, T]; L^r(\RR^3))$ for some $r$  satisfying  $r>3/2$ and $r\ge \gamma$, ${\bf m}\in  L^{\infty}([0, T]; L^s(\RR^3))$  ($s>3$), $(\eta, {\bf q})\in L^{\infty}([0, T]; L^1(\RR^3))$, where $\eta$ and  ${\bf q}$ are given in (4.3).  Moreover, we assume that $(\r, {\bf m})$ has the following additional regularity:
   \be\label{ca} \lim_{h\to 0}\int_0^t\int_{\RR^3}|\r(x, \tau+h)-\r(x, \tau)|dxd\tau=0, \quad t\in (0, T), a.e.\ee Then \\
    \be\label{6.2} E(t)\le E(0), \qquad 0<t<T,\ee
 and if $\gamma>\f{4}{3}$, then
 \begin{equation}\label{estimate1}
 H(t)\le C_1 H(0)+ C_2, \qquad  0<t<T,
 \end{equation}
 where $C_1$ and $C_2$ are two positive constants only depending on $\g$ and $M$ (cf. (\ref{6.1})), where
 $$H(t)= \int_{\RR^3}\{\f{\r^\g}{\g-1}+\f{|{\bf  m}|^2}{2\r}+\f{1}{8\pi}|\nabla\Phi|^2\}(x, t)dx, \quad t\in [0, T)$$.
  \end{thm}
 \begin{rem} (\ref{6.1}) and (\ref{6.2}) justify some assumptions made in Section 3 on the conservation of total mass and non-increase of energy. \end{rem}
 \begin{rem} The boundedness of $\int_{\RR^3}\r^\g(x,t) dx$ was proved in \cite{LY} for smooth solutions if $\g>4/3$. Here we prove that  this is still  true
 for general week solutions satisfying the entropy condition even without assuming that $\r\in L^{\infty}$. In fact, the global existence of radial  $L^{\infty}$-solutions was proved in \cite{Wang} for (\ref{1.1}) outside a ball. The blow-up of $L^{\infty}$-norm of the radial solutions of
 (\ref{1.1}) in the entire $\RR^3$ space was discussed in \cite{MK} and \cite{DXY}, respectively. \end{rem}
 \begin{rem}\label{rem12} Condition (\ref{ca}) can be assured by the following condition
 \be\label{cb} \lim_{\epsilon\to 0}\sup_{0\le \tau\le T, |y|\le 1}\int_{\RR^3}|\r(x, \tau)-\r(x-\epsilon y, \tau)dx=0, \ee
 if $(\r, {\bf m})\in L^{\infty}([0, T]; L^1(\RR^3))$; this is proved in the Appendix. Note that (\ref{ca}) is the $L^1$ modulus of continuity in time and
 (\ref{cb})  is the $L^1$ modulus of continuity in space. \end{rem}

In order to prove this theorem, we begin with the following lemma.
\begin{lem} If $f\in L^{r}(\RR^3)$ ($r\ge 1$), then
\be\label{0751}Bf\in \begin{cases}L^p(\RR^3), {\rm ~with~} 1/p=1/r-2/3, \qquad {\rm if~} r<3/2,\\
L^{\infty}(\RR^3), \qquad  {\rm if~} r\ge 3/2;\end{cases}\ee
and
\be\label{0752}\nabla (Bf)\in \begin{cases} L^q(\RR^3), {\rm ~with~} 1/q=1/r-1/3, \qquad {\rm if~} r<3,\\
L^{\infty}(\RR^3), \qquad   {\rm if~} r\ge 3.\end{cases}\ee
\end{lem}
The proof of this lemma follows from the extended Young inequality (cf. \cite{RS}, p. 32).
\begin{lem} Suppose $0\le \r\in L^{\infty}([0, T]; L^1(\RR^3))$  and
$\f{{\bf m}}{\sqrt\r}\in L^{\infty}([0, T]; L^2(\RR^3))$, then
\be\label{0753} {\bf m}\in L^{\infty}([0, T]; L^1(\RR^3)).\ee\end{lem}
\begin{proof} Using H$\ddot{\rm o}$lder inequality, we have
\be\label{0754} \int |{\bf m}|dx=\int \sqrt {\r}\f{|{\bf m}|}{\sqrt {\r}}dx\le (\int \r dx)^{1/2}(\int \f{|{\bf m}|^2}{\r})^{1/2}.\ee
Note that (\ref{0754}) implies ${\bf m}\in L^{\infty}([0, T]; L^1(\RR^3))$.
\end{proof}
\begin{rem} $\eta\in L^{\infty}([0, T]; L^1(\RR^3))$ implies $\f{{\bf m}}{\sqrt\r}\in L^{\infty}([0, T]; L^2(\RR^3))$.\end{rem}
\begin{lem} Let $(\rho, {\bf m}, \Phi)$ be a weak solution defined in Definition 3.1.  Suppose $(\rho, {\bf m}, \Phi)$ satisfies the entropy condition (4.2), $\rho\in  L^{\infty}([0, T];  L^1(\RR^3))\cap L^{\infty}([0, T]; L^r(\RR^3))$ for some $r$  satisfying  $r>3/2$ and $r\ge \gamma$, ${\bf m}\in  L^{\infty}([0, T]; L^s(\RR^3))$  ($s>3$), $(\eta, {\bf q})\in L^{\infty}([0, T]; L^1(\RR^3))$, where $\eta$ and  ${\bf q}$ are given in (4.3).
 Then, for any
$\tau\in [0, T)$, we have
\be\label{entropy12} \int_{\RR^3}\eta(x, \tau)dx-\int_0^{\tau}\int_{\RR^3} {\bf m}\cdot \nabla \Phi dxdt\le \int_{\RR^3}\eta(x, 0)dx, \quad  \tau\in (0, T), a.e. \ee
\end{lem}
\begin{proof}
For a fixed $\tau\in (0, T)$,  and small positive $\epsilon$ and $R>0$, we define
\be\label{thetat} \theta(t)=\begin{cases} & 1, \qquad\qquad\qquad 0\le t\le \tau,\\
&-\f{1}{\epsilon}(t-\tau)+1, \qquad \tau\le t\le \tau+\epsilon, \\
& 0, \qquad\qquad\qquad \tau+\epsilon\le t\le T,
\end{cases}\ee
and  for $x\in \RR^3$,
\be\label{alphax} \alpha(x)=\begin{cases} & 1, \qquad\qquad\qquad  |x|\le R,\\
&-\f{1}{\epsilon}(|x|-R)+1, \qquad R\le |x|\le R+\epsilon, \\
& 0, \qquad\qquad\qquad |x|\ge R+\epsilon.
\end{cases}\ee
Let $\beta(x, t)=\theta(t)\alpha(x)$, then $\beta(x, t)$ is Lipschitz continuous, with compact support in $[0, T)\times \RR^3$. Using
(\ref{entro}),  a calculation yields
\begin{align}\label{entropyj}
&-\frac{1}{\epsilon}\int_{\tau}^{\tau+\epsilon}\int_{|x|\le R}\eta(x, t)dxdt-\f{1}{\epsilon}\int_{\tau}^{\tau+\epsilon}\int_{R\le |x|\le R+\epsilon}
\eta(x, t)\alpha(x)dxdt\notag\\
&-\frac{1}{\epsilon}\int_0^{\tau+\epsilon}\int_{R\le |x|\le R+\epsilon}(\sum_{j=1}^3q_j\f{x_j}{|x|})\theta(t)dxdt\notag\\
&+\int_{|x|\le R}\eta(x,0)dx+\int _{R\le |x|\le R+\epsilon}\eta(x, 0) \alpha(x)dx\notag\\
&+\int_0^{\tau}\int_{|x|\le R} {\bf m}\cdot \nabla \Phi dxdt+\int_{\tau}^{\tau+\epsilon}\int_{|x|\le R+\epsilon}{\bf m}\cdot\nabla \Phi \beta(x, t)dxdt\ge 0.
\end{align}
Since $(\eta, {\bf q})\in L^{\infty}([0, T], L^1(\RR^3)$,
we have
\be\label{rxyz1}
\lim_{R\to \infty}\int_{R\le |x|\le R+\epsilon}\eta(x, t)\alpha(x)dx=0, \qquad a.e.,\ t\in [0, T], \ee
\be\label{rxyz2}
\lim_{R\to \infty}\int_{R\le |x|\le R+\epsilon}\eta(x, 0)\alpha(x)dx=0,  \ee
and
\be\label{rxyz3}
\lim_{R\to \infty}\int_{R\le |x|\le R+\epsilon}(\sum_{j=1}^3 q_j\f{x_j}{|x|})\theta(t) dx=0, \qquad a.e.,\  t\in [0, T]. \ee
We let $R\to \infty$ in (\ref{entropyj}) to get
\begin{align}\label{entropyk}
&-\frac{1}{\epsilon}\int_{\tau}^{\tau+\epsilon}\int_{\RR^3}\eta(x, t)dxdt
+\int_{\RR^3}\eta(x,0)dx\notag\\
&+\int_0^{\tau}\int_{\RR^3} {\bf m}\cdot \nabla \Phi dxdt+\int_{\tau}^{\tau+\epsilon}\int_{\RR^3}{\bf m}\cdot\nabla \Phi \beta(x, t)dxdt\ge 0.
\end{align}
Because $\r\in L^{\infty}([0, T]; L^r(\RR^3))$ with $r>3/2$,  by (5.7) we have $\nabla\Phi\in L^{\infty}([0, T]; L^q(\RR^3)$ with $q>3$. It then follows from (5.8),  the assumption
${\bf m}\in  L^{\infty}([0, T]; L^s(\RR^3))$ with $s>3$ and Holder's inequality that
$${\bf m}\cdot \nabla \Phi\in  L^{\infty}([0, T]; L^1(\RR^3)).$$
This implies
$$\lim_{\epsilon\to 0}\int_{\tau}^{\tau+\epsilon}\int_{\RR^3}{\bf m}\cdot\nabla \Phi \beta(x, t)dxdt=0.$$
Letting $\epsilon\to 0$ in (\ref{entropyk}), we obtain (\ref{entropy12}). \end{proof}

\begin{lem}\label{phit} Let $(\rho, {\bf m}, \Phi)$ be an entropy  weak solution defined in Section 4  satisfying the conditions in Lemma 5.3. Then
\be\label{171} \pa_t\Phi(x, t)=-\int_{\RR^3}{\bf m}(y, t)\cdot \nabla_y(\f{1}{|y-x|})dy.\ee
Moreover
\be\label{172} \pa_t\Phi\in  L^{\infty}([0, T]; L^1(\RR^3)),\ee
and
\be\label{173} \pa_t \Phi\in
L^{\infty}([0, T]\times\RR^3). \ee\end{lem}
\begin{proof} The key is to prove (\ref{171}). Once (\ref{171}) is proved, (\ref{172}) and (\ref{173}) follow from the fact that
${\bf m}\in L^{\infty}([0, T]; L^1(\RR^3))\cap L^{\infty}([0, T]; L^s(\RR^3))$ and the extended Young's inequality (cf. \cite{RS}, p.32).
In order to prove (\ref{171}), we use the fact that $(\r, {\bf m})$ satisfies the first equation of (\ref{1.1}) in the sense of distributions.
For this purpose,  we choose a $C^{\infty}$ function  $\delta(z) $ ($z\in \RR^1$) with compact support in the interval $[1, 2]$ satisfying $0\le \delta(z)\le 1$ and $\int_{-\infty}^{+\infty}\delta(z)dz=1$, and let
 \be\label{174} \delta_{\epsilon}(z)=\f{1}{\epsilon}\delta(\f{z}{\epsilon}),\  \alpha_{\epsilon}(z)=\int_{-\infty}^z \delta_{\epsilon}(s)ds,
 \quad z\in \RR^1,
 \ee
 for small positive $\epsilon$. For $y\in \RR^3$, $0<\epsilon<\f{1}{2}$ and $R>1$, we set
 \be f_{\epsilon}^R(y)=\begin{cases} &\alpha_{\epsilon}(|y|),  \qquad\qquad\qquad  |y|\le \frac{R}{2}+\epsilon,\\
 & \alpha_{\epsilon}(R+2\epsilon-|y|), \qquad  |y|\ge \f{R}{2}+\epsilon.\end{cases}
 \ee
 Then
 \be \begin{cases}  & f_{\epsilon}^R(y)=0,\qquad\qquad as\ |y|\le \epsilon,\ or\ |y|\ge R+\epsilon, \\
 & 0\le f_{\epsilon}^R(y)\le 1, \qquad as\ \epsilon\le |y|\le 2\epsilon, \ or\ R\le |y|\le R+\epsilon,\\
 & f_{\epsilon}^R(y)=1,\qquad\qquad as\ 2\epsilon\le |y|\le R.\end{cases}\ee
 For $x\in \RR^3$,  we choose
 \be g_{\epsilon}^R(y)=f_{\epsilon}^R(y-x)\frac{1}{|y-x|}.\ee
 Then $g_{\epsilon}^R(y)\in  C^{\infty}_0(\RR^3)$ for any fixed $x\in \RR^3.$
 Since $(\r, {\bf m})$ satisfies the first equation of  (\ref{1.1}) in the sense of distributions, it is easy to
 show (see \cite{freisel} for instance), $\int_{\RR^3}\r(y, t) g_{\epsilon}^R(y)dy$ is differentiable in $t$ for
 $t\in [0, T]$ a. e.,  and satisfies
 \be\label{keyidea}
 \f{d}{dt}\int_{\RR^3}\r(y, t) g_{\epsilon}^R(y)dy=\int_{\RR^3}{\bf m}(y, t)\cdot \nabla_yg_{\epsilon}^R(y)dy, \qquad t\in [0, T],\ a.~e.\ee
 We also let
 \be\label{gepsilon} g_{\epsilon}(y)=\lim_{R\to \infty} g_{\epsilon}^R(y), \qquad y\in\RR^3. \ee
 Then we show (\ref{171}) in the following steps.\\
 \underline{Step 1}. We show that $\int_{\RR^3}\r(y, t)g_{\epsilon}(y)dy$ is differentiable for $t\in (0, T]$,  a.e.,  and
 \be\label{step1}\f{d}{dt}\int_{\RR^3}\r(y, t)g_{\epsilon}(y)dy=\int_{\RR^3}{\bf m}\cdot \nabla_y g_{\epsilon}(y)dy,\ee
 for $t\in (0, T], a. e.$ \\
  For this purpose, we prove that
 \begin{align}\label{step11}& \f{1}{h} \int_{\RR^3}\f{\r(y, t+h)-\r(y, t)}{h}g_{\epsilon}^R(y)dy \to
 \int_{\RR^3}{\bf m}(y, t)\cdot \nabla_yg_{\epsilon}^R(y)dy\notag\\&{\rm as} ~{h\to 0}
 ~{\rm uniformly~ in}~ R ~{\rm for} ~R\ge 1.\end{align}
 This is proved as follows. Since $(\r, {\bf m})$ satisfies the first equation of (\ref{1.1}) in the sense of distributions and
 $g_{\epsilon}^R(y)\in C^{\infty}_c(\RR^3)$, it is easy to verify (see \cite{freisel} for instance),
 \be\label{tt}\int_{\RR^3}(\r(y, t+h)-\r(y, t))g_{\epsilon}^R(y)dy=\int_t^{t+h}\int_{\RR^3}{\bf m}(y,s)\cdot \nabla_yg_{\epsilon}^R(y)dyds, \ee
 for $[t, t+h]\subset [0, T].$ Thus,
 \be\label{tt1}\lim_{h\to 0}\int_{\RR^3}\f{(\r(y, t+h)-\r(y, t))}{h}g_{\epsilon}^R(y)dy=\lim_{h\to 0}\f{1}{h}\int_t^{t+h}\int_{\RR^3}{\bf m}(y,s)\cdot \nabla_yg_{\epsilon}^R(y)dyds. \ee On the other hand,
\begin{align}\label{geg}
&\f{1}{h}\int_t^{t+h}\int_{\RR^3}{\bf m}(y,s)\cdot \nabla_yg_{\epsilon}^R(y)dyds-\int_{\RR^3}{\bf m}(y,t)\cdot \nabla_yg_{\epsilon}^R(y)dy\notag\\
&=\f{1}{h}\int_t^{t+h}\int_{\RR^3}({\bf m}(y,s)-{\bf m}(y,t))\cdot \nabla_yg_{\epsilon}^R(y)dyds\notag\\
&=\f{1}{h}\int_t^{t+h}\int_{\epsilon\le |y-x|\le R}({\bf m}(y,s)-{\bf m}(y,t))\cdot \nabla_y(\f{\alpha_{\epsilon}(|y-x|)}{|y-x|})dyds\notag\\
&+\f{1}{h}\int_t^{t+h}\int_{R\le |y-x|\le R+\epsilon}({\bf m}(y,s)-{\bf m}(y,t))\cdot \nabla_yg_{\epsilon}^R(y)dyds.\end{align}
The first term can be handled as follows. For $h>0$,
\begin{align}\label{gege1}
&|\f{1}{h}\int_t^{t+h}\int_{\epsilon\le |y-x|\le R}({\bf m}(y,s)-{\bf m}(y,t))\cdot \nabla_y(\f{\alpha_{\epsilon}(|y-x|)}{|y-x|})dyds|\notag\\
&\le |\f{1}{h}\int_t^{t+h}\int_{\epsilon\le |y-x|\le R}|{\bf m}(y,s)-{\bf m}(y,t)|(\f{\delta_{\epsilon}(|y-x|)}{|y-x|}+\f{\alpha_{\epsilon}(|y-x|)}{|y-x|^2})dyds\notag\\
&\le \f{2}{\epsilon h}\int_t^{t+h}\int_{\epsilon\le |y-x|\le 2\epsilon}|{\bf m}(y,s)-{\bf m}(y,t)|\f{1}{|y-x|}dyds\notag\\
&+ \f{1}{ h}\int_t^{t+h}\int_{2\epsilon\le |y-x|\le R}|{\bf m}(y,s)-{\bf m}(y,t)|\f{1}{|y-x|^2}dyds\notag\\
&\le \f{2}{\epsilon^2 h}\int_t^{t+h}\int_{\epsilon\le |y-x|\le 2\epsilon}|{\bf m}(y,s)-{\bf m}(y,t)|dyds\notag\\
&+ \f{1}{ 4\epsilon^2 h}\int_t^{t+h}\int_{2\epsilon\le |y-x|\le R}|{\bf m}(y,s)-{\bf m}(y,t)|dyds.\end{align}
The last term in (\ref{geg}) can be estimated as follows.
\begin{align}\label{gege2}
&|\f{1}{h}\int_t^{t+h}\int_{R\le |y-x|\le R+\epsilon}({\bf m}(y,s)-{\bf m}(y,t))\cdot \nabla_yg_{\epsilon}^R(y)dyds|\notag\\
&\le (\f{1}{\epsilon }+\frac{1}{R})\f{1}{h}\int_t^{t+h}\int_{R\le |y-x|\le R+\epsilon}|{\bf m}(y,s)-{\bf m}(y,t)|\f{1}{|y-x|}dyds\notag\\
&\le  (\f{1}{\epsilon }+\frac{1}{R})\f{1}{h} \int_t^{t+h}\int_{\RR^3}|{\bf m}(y,s)-{\bf m}(y,t)|dyds.\end{align}
Since we choose $R>1$, (\ref{geg}), (\ref{gege1}) and (\ref{gege2}) yield,
\begin{align}\label{geg1}
&|\f{1}{h}\int_t^{t+h}\int_{\RR^3}{\bf m}(y,s)\cdot \nabla_yg_{\epsilon}^R(y)dyds-\int_{\RR^3}{\bf m}(y,t)\cdot \nabla_yg_{\epsilon}^R(y)dy|\notag\\
&\le (\f{9}{4\epsilon^2}+\f{1}{\epsilon}+1)\f{1}{h}\int_t^{t+h}\int_{\RR^3}|{\bf m}(y,s)-{\bf m}(y,t)|dyds.\end{align}

Since ${\bf m}\in L^{\infty}([0, T]; L^1(\RR^3)$, we have
\be \lim_{h\to 0+}\f{1}{  h}\int_t^{t+h}\int_{\RR^3}|{\bf m}(y,s)-{\bf m}(y,t)|dyds=0, \quad t\in [0, T],\  a. e.\ee
Therefore $\f{1}{h}\int_t^{t+h}\int_{\RR^3}({\bf m}(y,s)-{\bf m}(y,t))\cdot \nabla_yg_{\epsilon}^R(y)dyds$ converges to
zero as $h\to 0+$ uniformly in $R$ for $R>1$.  By a similar approach, we can show that  $\f{1}{h}\int_{t-h}^{t}\int_{\RR^3}({\bf m}(y,s)-{\bf m}(y,t))\cdot \nabla_yg_{\epsilon}^R(y)dyds$ converges to zero as $h\to 0-$ uniformly in $R$ for $R>1$. This verifies (\ref{step11}). (\ref{step1})  follows by the following
argument, using (5.22) and (5.24).
 \begin{align}\label{step12}&\f{d}{dt}\int_{\RR^3}\r(y, t)g_{\epsilon}(y)dy\notag\\
& =\lim_{h\to 0}\lim_{R\to \infty}\int_{\RR^3}\f{\r(y, t+h)-\r(y, t)}{h}g_{\epsilon}^R(y)dy\notag\\
 &= \lim_{R\to \infty}\lim_{h\to 0}\int_{\RR^3}\f{\r(y, t+h)-\r(y, t)}{h}g_{\epsilon}^R(y)dy\notag\\
 &=\lim_{R\to \infty}\f{d}{dt}\int_{\RR^3}\r(y, t)g_{\epsilon}^R(y)dy\notag\\
 &=\lim_{R\to \infty}\int_{\RR^3}{\bf m}\cdot \nabla_y g_{\epsilon}^R(y)dy\notag\\
 &=\int_{\RR^3}{\bf m}\cdot \nabla_y g_{\epsilon}(y)dy.\end{align}
 \underline{Step 2.}  In this step, we show that
 \begin{align}\label{71211}
& \int_{\RR^3}\r(y, t)g_{\epsilon}(y)dy \to \int_{\RR^3}\f{\r(y, t)}{|y-x|}dy{\rm~ as~}\epsilon\to 0\notag\\
&{\rm~uniformly~in~} t  {\rm~ for~} t\in (0, T),\end{align}
and
\be\label{71212}\lim_{\epsilon\to 0}\int_{\RR^3}{\bf m}(y,t)\cdot\nabla_yg_{\epsilon}(y)dy=\int_{\RR^3}{\bf m}(y,t)\cdot\nabla_y(\f{1}{|y-x|})dy,
\ee for $t\in (0, T)$. \\
We prove (\ref{71211}) as follows. Since $\r\in L^{\infty}([0, T]; L^r(\RR^3))$ with $r>3/2$ and $r\ge \gamma$,  we have, by using H$\rm\ddot{o}$lder inequality,
\begin{align}\label{71213}
&|\int_{\RR^3}\f{\r(y, t)}{|y-x|}dy-\int_{\RR^3}\r(y, t)g_{\epsilon}(y)dy|\le \int_{\epsilon\le |y-x|\le 2\epsilon}\f{\r(y, t)}{|y-x|}dy\notag\\
&\le (\int_{\epsilon\le |y-x|\le 2\epsilon}\r^r(y, t)dy)^{1/r}(\int_{\epsilon\le |y-x|\le 2\epsilon}\f{1}{|y-x|^l}dy)^{1/l}\notag\\
&\le ||\r||_{L^{\infty}([0, T]; L^r(\RR^3))}{(\int_{\epsilon}^{2\epsilon} 4\pi s^{2-l}ds)}^{1/l},
\end{align}
where $l=\f{r}{r-1}$. Since $r>3/ 2$, $l<3$, (\ref{71211}) follows.  Next (\ref{71212}) can be shown as follows.
Since ${\bf m}\in L^{\infty}([0, T]; L^s(\RR^3))$ for $s>3$, we have
\begin{align}\label{71214}
&|\int_{\RR^3}{\bf m}(y, t)\cdot\nabla_yg_{\epsilon}(y)dy-\int_{\RR^3}{\bf m}(y, t)\cdot\nabla_y\f{1}{|y-x|}dy|\notag\\
&=|\int_{\epsilon\le |y-x|\le 2\epsilon} {\bf m}(y, t)\cdot \nabla_y(g_{\epsilon}(y)-\f{1}{|y-x|})dy|\notag\\
&=|\int_{\epsilon\le |y-x|\le 2\epsilon} {\bf m}(y, t)\cdot \nabla_y(\f{1}{|y-x|}(\alpha_{\epsilon}(|y-x|)-1))dy|\notag\\
&\le\int_{\epsilon\le |y-x|\le 2\epsilon} |{\bf m}(y, t)|(\f{1}{|y-x|^2}+\f{1}{|y-x|}\delta_{\epsilon}(|y-x|)))dy\notag\\
&\le\f{2}{\epsilon}\int_{\epsilon\le |y-x|\le 2\epsilon} |{\bf m}(y, t)|\f{1}{|y-x|}dy\notag\\
&\le \f{2}{\epsilon}||{\bf m}||_{L^{\infty}([0, T]; L^s(\RR^3))}(\int_{\epsilon\le |y-x|\le 2\epsilon} \f{1}{|y-x|^q}dy)^{1/q}\notag\\
&\le \f{2}{\epsilon}||{\bf m}||_{L^{\infty}([0, T]; L^s(\RR^3))}\left(\int_{\epsilon}^{ 2\epsilon}4\pi \tau^{2-s'}d\tau \right)^{1/s'},
\end{align}
where $q=\f{s}{s-1}$. Since $s>3$, then $q<3/2$. Therefore, (\ref{71212}) is proved. \\
By (\ref{71211}) and (5.24), we have that $\int_{\RR^3}\f{\r(y, t)}{|y-x|}dy$ is differentiable with respect to $t$ for $(t, x)\in (0 , T)\times \RR^3$, a. e.
Moreover, by (5.24), (\ref{71211}) and (\ref{71212}), we obtain,
\begin{align}
\f{d}{dt}\int_{\RR^3}\f{\r(y, t)}{|y-x|}dy&=\f{d}{dt}(\lim_{\epsilon\to 0}\int_{\RR^3}\r(y, t)g_{\epsilon}(y)dy)\notag\\
&=\lim_{\epsilon\to 0}\f{d}{dt}\int_{\RR^3}\r(y, t)g_{\epsilon}(y)dy=\lim_{\epsilon\to 0}\int_{\RR^3}{\bf m}(y, t)\cdot \nabla_yg_{\epsilon}(y)dy\notag\\
&=\int_{\RR^3}{\bf m}(y, t)\cdot \nabla_y(\f{1}{|y-x|})dy. \end{align}
This proves (5.18).  (5.19) and (5.20) then follows as we showed at the beginning of the proof of this Lemma.
  \end{proof}

 {\it Proof of  Theorem 5.1}\\
 We prove Theorem 5.1 in the following steps. \\
 \vskip 0.2cm
\noindent \underline{Step 1}  In this step, we prove (5.1). This can be proved by using  (5.25) in which $g_{\epsilon}^R(y)$ is replaced by $f_{\epsilon}^R(y)$,
  i.e.,
 \be\label{5.441}\f{d}{dt}\int_{\RR^3}\r(y, t)f_{\epsilon}^R(y)dy=\int_{\RR^3}{\bf m}(y, t)\cdot \nabla_y f_{\epsilon}^R(y)dy, t\in [0, T], a. e., \ee
 where $f_{\epsilon}^R$ is defined in (5.22). We integrate ({\ref{5.441}) to get
  \be\label{5.442}\int_{\RR^3}\r(y, t)f_{\epsilon}^R(y)dy-\int_{\RR^3}\r(y, 0)f_{\epsilon}^R(y)dy=\int_0^t\int_{\RR^3}{\bf m}(y,s)\cdot \nabla_y f_{\epsilon}^R(y)dy,  \ee
  By using a same argument as in the proof of Lemma 5.4, we can prove
  $$\lim_{\epsilon\to 0}\lim_{R\to \infty}\int_{\RR^3}\r(y, t)f_{\epsilon}^R(y)dy=\int_{\RR^3}\r(y, t)dy,$$ $$\lim_{\epsilon\to 0}\lim_{R\to \infty}\int_{\RR^3}\r(y, 0)f_{\epsilon}^R(y)dy=\int_{\RR^3}\r(y, 0)dy, $$
  and
  $$\lim_{\epsilon\to 0}\lim_{R\to \infty}\int_0^t\int_{\RR^3}{\bf m}(y,s)\cdot \nabla_y f_{\epsilon}^R(y)dy=0.$$
  (5.1) follows from (\ref{5.442}) by letting $R\to \infty$ and $\epsilon\to 0$.
  \vskip 0.2cm

\noindent  \underline{Step 2} In this step, we show that
 \be\label{5.41}
 \int_0^t\int_{\RR^3} \rho(x, s)\pa_s \Phi(x, s)dxds=\f{1}{2}\left(\int_{\RR^3}(\r\Phi)(x, t)dx-\int_{\RR^3}(\rho\Phi)(x, 0)dx\right), \  t\in [0, T).  \ee
 This is can be proved as follows.
 \begin{align}\label{5.42}
& \int_0^t \int_{\RR^3}\rho(x, s)\pa_s \Phi(x, s)dxds\notag\\
&=\lim_{
h\to 0}\f{1}{h}\int_0^t\int_{\RR^3}\r(x, s)\int_{\RR^3}\f{\r(y, s+h)-\r(y, s)}{|x-y|}dydxds\notag\\
&= \lim_{h\to 0}\f{1}{h}\left(\int_h^{t+h}\int_{\RR^3}\int_{\RR^3}\f{\r(x, s-h)\r(y, s)}{|x-y|}dydxds-\int_0^t\int_{\RR^3}\int_{\RR^3}\f{\r(x, s)\r(y, s)}{|x-y|}dydxds\right)\notag\\
&=\lim_{h\to 0}\f{1}{h}\int_h^{t}\int_{\RR^3}\int_{\RR^3}\f{(\r(x, s-h)-\r(x, s))\r(y, s)}{|x-y|}dydxds\notag\\
&+\lim_{h\to 0}\f{1}{h}\int_t^{t+h}\int_{\RR^3}\int_{\RR^3}\f{\r(x, s-h)\r(y, s)}{|x-y|}dydxds
-\lim_{h\to 0}\f{1}{h}\int_0^{h}\int_{\RR^3}\int_{\RR^3}\f{\r(x, s-h)\r(y, s)}{|x-y|}dydxds\notag\\
&=\lim_{h\to 0}\f{1}{h}\int_h^{t}\int_{\RR^3}\int_{\RR^3}\f{(\r(x, s-h)-\r(x, s))\r(y, s)}{|x-y|}dydxds\notag\\
&+\int_{\RR^3}(\r \Phi)(x, t)dx-\int_{\RR^3}(\r\Phi)(x, 0)dx.  \end{align}
On the other hand,
 \begin{align}\label{5.43}
&\f{1}{h}\int_h^{t}\int_{\RR^3}\int_{\RR^3}\f{(\r(x, s-h)-\r(x, s))\r(y, s)}{|x-y|}dydxds\notag\\
&=\f{1}{h}\int_0^{t}\int_{\RR^3}\int_{\RR^3}\f{(\r(x, \tau)-\r(x, \tau+h))\r(y, \tau)}{|x-y|}dydxd\tau\notag\\
&+\f{1}{h}\int_0^{t}\int_{\RR^3}\int_{\RR^3}\f{(\r(x, \tau)-\r(x, \tau+h))(\r(y, \tau+h)-\rho(y, \tau))}{|x-y|}dydxd\tau\notag\\
&-\f{1}{h}\int_{t-h}^{t}\int_{\RR^3}\int_{\RR^3}\f{(\r(x, \tau)-\r(x, \tau+h))\r(y, \tau+h)}{|x-y|}dydxd\tau.\end{align}
Since $$\lim_{h\to 0}\f{1}{h}\int_{\RR^3}\f{\r(y, \tau+h)-\r(y, \tau)}{|x-y|}dy=-\pa_{\tau}\Phi(x, \tau)$$,
\be\label{5.44}|\f{1}{h}\int_{\RR^3}\f{\r(y, \tau+h)-\r(y, \tau)}{|x-y|}dy|\le |\pa_{\tau}\Phi(x, \tau)|+1,\ee
for small $|h|$. Therefore,
\begin{align}\label{5.45}&|\f{1}{h}\int_0^{t}\int_{\RR^3}\int_{\RR^3}\f{(\r(x, \tau)-\r(x, \tau+h))(\r(y, \tau+h)-\rho(y, \tau))}{|x-y|}dydxd\tau |\notag\\
&\le (||\pa_t\Phi ||_{L^{\infty}([0, T]\times \RR^3}+1)\int_0^{t}\int_{\RR^3}|(\r(y, \tau+h)-\r(x, \tau)|dyd\tau.
\end{align}
Then (\ref{ca}), (5.20) and (\ref{5.45}) imply
\be\label{5.46} \lim_{h\to 0}\f{1}{h}\int_0^{t}\int_{\RR^3}\int_{\RR^3}\f{(\r(x, \tau)-\r(x, \tau+h))(\r(y, \tau+h)-\rho(y, \tau))}{|x-y|}dydxd\tau=0.\ee
Similarly, we have, for small $|h|$,
\begin{align}\label{5.47}
&|\f{1}{h}\int_{t-h}^{t}\int_{\RR^3}\int_{\RR^3}\f{(\r(x, \tau)-\r(x, \tau+h))\r(y, \tau+h)}{|x-y|}dydxd\tau|\notag\\
&\le (||\pa_t\Phi ||_{L^{\infty}([0, T]\times \RR^3}+1)\int_{t-h}^{t}\int_{\RR^3}\r(y, \tau+h)dyd\tau.
\end{align}
Since $\rho\in L^{\infty}([0, T]; L^1(\RR^3))$, (\ref{5.47}) implies,
\be\label{5.48}\lim_{h\to 0}\f{1}{h}\int_{t-h}^{t}\int_{\RR^3}\int_{\RR^3}\f{(\r(x, \tau)-\r(x, \tau+h))\r(y, \tau+h)}{|x-y|}dydxd\tau=0.
\ee
Hence, (\ref{5.43}), (\ref{5.46}) and (\ref{5.48}) yield
\begin{align}\label{5.49}
&\lim_{h\to 0}\f{1}{h}\int_h^{t}\int_{\RR^3}\int_{\RR^3}\f{(\r(x, s-h)-\r(x, s))\r(y, s)}{|x-y|}dydxd\tau\notag\\
&=\lim_{h\to 0}\f{1}{h}\int_0^{t}\int_{\RR^3}\int_{\RR^3}\f{(\r(x, \tau)-\r(x, \tau+h))\r(y, \tau)}{|x-y|}dydxd\tau.
\end{align}
This, together with (\ref{5.42}), implies (\ref{5.41}).
\vskip 0.2cm
\noindent\underline{Step 3} In this step, we prove (5.3). \\
Since $\r\in L^{\infty}([0, T; L^1(\RR^3))\cap L^{\infty}([0, T]; L^r(\RR^3))$, where $r>3/2$ and $r\ge \gamma$, we have, in view of (5.7) that
\be\label{5.52} \nabla \Phi\in L^{\infty} ([0, T]; L^{3/2}(\RR^3))\cap L^{\infty}([0, T]; L^{\lambda}(\RR^3)), \ee
if $r<3$, where $\f{1}{\lambda}=\f{1}{r}-\f{1}{3}$. We also know that $\lambda>3$ if $r>3/2$. Similarly, by (5.7), we have
\be\label{5.53} \nabla \Phi\in L^{\infty} ([0, T]; L^{3/2}(\RR^3))\cap L^{\infty}([0, T]\times \RR^3 ), \ee
if $r\ge 3$. Furthermore, because $(\r, {\bf m})$ satisfies the first equation of (1.1) in the sense of distributions, then by a density
argument as in \cite{freisel}, in view of (5.19),  (5.20), (\ref{5.52} and (\ref{5.53}), we have,
\begin{align}\label{5.54}
&\int_{\RR^3}(\r\Phi)(x, t)dx-\int_{\RR^3}(\r\Phi)(x, 0)dx\notag\\
&=\int_0^t\int_{\RR^3}\rho(x, s)\pa_s\Phi(x,s)dxds+\int_0^t\int_{\RR^3}{\bf m}(x, s)\cdot \nabla\Phi(x,s)dxds, \end{align}
for $t\in [0, T)$. This, together with (5.10) and (5.44), implies (5.3), due to the fact
\be\label{5.55} E(t)=\int_{\RR^3}\eta(x, t)dx-\f{1}{2}\int_{\RR^3}(\r\Phi)(x, t)dx, \ee
for $t\in [0, T)$.
\vskip 0.2cm

\noindent\underline{Step 4} In this step, we proof (5.4).\\
First, since $\r\in L^{\infty}([0, T]; L^1(\RR^3)\cap  L^{\infty}([0, T]; L^r(\RR^3)$ with $r>3/2$, it follows from \cite{lieb}), \cite{rein1} and \cite{Rein} that
$$\f{1}{2}\int_{\RR^3}(\r\Phi)(x, t)dx= -\f{1}{8\pi}\int_{\RR^3}|\nabla \Phi|^2(x, t)dx, \quad t\in [0, T].$$
Using (2.19), we have, for $\g>4/3$
\be\label{00x}\f{1}{8\pi}|\nabla\Phi|^2\}dx=\int
  \frac{1}{2}\rho B\rho dx
  \le
  C\int \r^{4/3}dx(\int \r dx)^{2/3}=M^{2/3}\int \r^{4/3}dx,
  \ee
  where $A(\r)$ is given by (2.3).
   Taking $p=1$, $q=4/3$, $r=\g$,  and $a=\f{\f{3}{4}\g-1}{\g-1}$ in Young's inequality (2.17), we obtain,
  \be ||\r||_{4/3}\le ||\r||_1^a||\r||_{\g}^{1-a}=M^a||\r||_{\g}^{1-a}.\ee
  This is
  \be\label{hahax}\int\r^{4/3}dx\le M^{\f{4}{3}a}(\int\r^\g dx)^b,
  \ee
  where $b=\f{1}{3(\g-1)}$. Since $\g>4/3$, we have $0<b<1$. Therefore, (\ref{00x}) and (\ref{hahax}) imply
  \be\la{0001x}  \int\frac{1}{2}\rho B\rho dx
\le C(\g-1)^bM^{\f{4}{3}a+\f{2}{3}}(\int A(\r)dx)^b.\ee
  Using the  inequality (cf.[15] p. 145)
  \be\la{0002x}\alpha\beta\le \epsilon\alpha^s+\epsilon^{-t/s}\beta^t,\ee
   if  $s^{-1}+t^{-1}=1$ ($s, t>1$) and $\epsilon>0$, since $b<1$,   we can bound  $C(\g-1)^bM^{\f{4}{3}a+\f{2}{3}}(\int A(\r)dx)^b$ by $\f{1}{2}\int A(\rho)dx+C_2$,
   where $C_2$ is a constant depending only on $M$ and $\g$ (we can take $\epsilon=1/2$ and $s=1/b$ and $t=(1-s^{-1})^{-1}$ in (\ref{0002}) since $s>1$ due to $0<b<1$). Therefore,
 \be\label{5.56} \f{1}{2}|\int_{\RR^3}(\r\Phi)(x,t)dx|=\f{1}{8\pi}||\nabla\Phi(\cdot, t)||_2^2\le \f{1}{2}\int_{\RR^3}\f{\r^\g(x, t)}{\g-1}dx+C,\ee
for $t\in [0, T),$ where $C$ is a constant only depending on $M=\int_{\RR^3}\int \r(x,t)dx=\int_{\RR^3}\r(x, 0)dx$ (cf. (5.1)) and $\g$.
 This, together with (5.3), implies (5.4).

\section{ Appendix}

In this appendix,  we prove the  following theorem which is  Remark \ref{rem12} in Section 5. \\

\noindent {\bf Theorem } If  $(\r, {\bf m})\in L^{\infty}([0, T]; L^1(\RR^3))$ satisfies the first equation of (\ref{1.1}) in the sense of distributions, then
\be\label{cb1} \lim_{\epsilon\to 0}\sup_{0\le t\le T, |y|\le 1}\int_{\RR^3}|\r(x, t)-\r(x-\epsilon y, t)|dx=0, \quad t\in (0, T), a.e., \ee
implies
   \be\label{ca1} \lim_{h\to 0}\int_0^t\int_{\RR^3}|\r(x, t+h)-\r(x, t)|dx=0, \quad t\in (0, T), a.e.\ee

\begin{proof} For any fixed $t \in (0, T)$ and small $h$, we let
$$w(x)=\r(x, t+h)-\rho(x, t).$$
First, we note that if $\psi(x)\in C^1(\RR^3))$ with $\psi$ and $\nabla\psi$ being bounded in $\RR^3$, then
\be\label{7.1} \int_{\RR^3}w(x)\psi(x)dx=\int_{t}^{t+h}\int_{\RR^3}{\bf m}(x, s)\cdot\nabla \psi(x)dxds.\ee
This is because $(\rho, {\bf m}) \in L^{\infty}([0, T]; L^1(\RR^3))$ satisfies the first equation of (\ref{1.1}) in the sense of distributions.
The justification of (\ref{7.1}) is standard, for instance, see \cite{freisel}. In view of (\ref{7.1}), we have
\be\label{7.2} |\int_{\RR^3}w(x)\psi(x)dx|\le h \sup_{x\in \RR^3}|\nabla\psi(x)|||m||_{L^{\infty}([0, T]; L^1(\RR^3))}.\ee
We choose $\psi$ as
$$\psi(x)=\int_{\RR^3}sgn (x-\epsilon y)\delta(y)dy, $$
where $sgn$ is the sign function, $\delta\in C_0^\infty(\RR^3)$ is a smooth function satisfying $0\le \delta(y)\le 1$,$\int_{\RR^3}\delta(y)dy=1$ and
$supp\ \delta\subset \{y\in \RR^3: |y|\le 1\}$. Then
$|\nabla\psi|\le \f{C}{\epsilon}$ for some constant $C$. Moreover,
\begin{align}\label{7.3}
&|\int_{\RR^3}w(x)\psi(x)dx-\int_{\RR^3}|w(x)|dx|\notag\\
&=|\int_{\RR^3}\int_{\RR^3}(w(x)-w(x-\epsilon y))sgn(x-\epsilon y)\delta(y)dydx|\notag\\
&\le \sup_{|y|\le 1}\int_{\RR^3}|w(x)-w(x-\epsilon y)|dy\notag\\
&\le \sup_{|y|\le 1}\int_{\RR^3}|\r(x, t)-\r(x-\epsilon y, t)|dx+\sup_{|y|\le 1}\int_{\RR^3}|\r(x, t+h)-\r(x-\epsilon y, t+h)|dx.
\end{align}
Therefore,
\begin{align}\label{7.4}
&\int_{\RR^3}|w(x)|dx\notag\\
& \le \sup_{|y|\le 1}\int_{\RR^3}|\r(x, t)-\r(x-\epsilon y, t)|dx+\sup_{|y|\le 1}\int_{\RR^3}|\r(x, t+h)-\r(x-\epsilon y, t+h)|dx\notag\\
&+\f{Ch}{\epsilon}||m||_{L^{\infty}([0, T]; L^1(\RR^3))}.\end{align}
We let $h\to 0$ first in (\ref{7.4}), (\ref{ca1}) follows from (\ref{cb1}) because $\epsilon$ is arbitrary.
\end{proof}

\centerline{\bf Acknowledgments}
Luo  was supported in part by  the National Science Foundation under Grant  DMS-0606853. Smoller  was supported in part by  the National Science Foundation under Grant  DMS-0603754. We  thank Wen-Qing Xu for a helpful discussion.
Luo is grateful to Cathleen Morawetz for her encouragement and interest in this work.

\bibliographystyle{plain}

 \newpage

Tao Luo\\
Department of Mathematical Sciences ,\\  Worcester Polytechnic Institute\\
100 Institute Road, Worcester, MA 01609-2280\\
E-mail: tl48@georgetown.edu\\

Joel Smoller\\
 Department of Mathematics,\\ University of
Michigan\\ 525 East University Ave. \\Ann Arbor, MI 48109-1109. \\
E-mail: smoller@umich.edu\\

\end{document}